\documentclass[twocolumn]{emulateapj}
\usepackage{graphicx,natbib,color,amsmath,subfigure}
\usepackage{ulem}
\newcommand{\blank}[1]{}
\usepackage{color}
\begin{document}
\title{Solar ALMA observations: constraining the chromosphere above sunspots}

\author{Maria A. Loukitcheva}
\affil{Center For Solar-Terrestrial Research, New Jersey Institute of Technology, Newark, NJ 07102, USA}
\affil{Max Planck Institute for Solar System Research, Justus-von-Liebig-Weg 3, D-37073 G\"{o}ttingen, Germany}
\affil{Saint Petersburg State University, 7/9 Universitetskaya nab., St. Petersburg 199034, Russia}



\author{Kazumasa Iwai}
\affiliation{Institute for Space-Earth Environmental Research, Nagoya University, Furo-cho, Chikusa-ku, Nagoya 464-8601, Japan}

\author{Sami K. Solanki}
\affiliation{Max Planck Institute for Solar System Research, Justus-von-Liebig-Weg 3, D-37073 G\"{o}ttingen, Germany}
\affiliation{School of Space Research, Kyung Hee University, Yongin, Gyeonggi 446-701, Korea}

\author{Stephen M. White}
\affiliation{Space Vehicles Directorate, Air Force Research Laboratory, Albuquerque, NM, USA}

\author{Masumi Shimojo}
\affiliation{National Astronomical Observatory of Japan (NAOJ), National Institutes of Natural Sciences (NINS), Mitaka, Tokyo, 181-8588, Japan}
\affiliation{Department of Astronomical Science, School of Physical Science, SOKENDAI (The Graduate University of Advanced Studies), Mitaka, Tokyo, 181-8588, Japan}

\begin{abstract}
We present the first high-resolution Atacama Large Millimeter/Submillimeter Array (ALMA) observations of a sunspot at wavelengths of 1.3~mm and 3~mm, obtained during the solar ALMA Science Verification campaign in 2015, and compare them with the predictions of semi-empirical sunspot umbral/penumbral atmosphere models. For the first time millimeter observations of sunspots have resolved umbral/penumbral brightness structure at the chromospheric heights, where the emission at these wavelengths is formed.
We find that the sunspot umbra exhibits a radically different appearance at 1.3~mm and 3~mm, whereas the penumbral brightness structure is similar at the two wavelengths. The inner part of the umbra is $\sim$600 K brighter than the surrounding quiet Sun (QS) at 3 mm and is $\sim$700 K cooler than the QS at 1.3 mm, being the coolest part of sunspot at this wavelength. On average, the brightness of the penumbra at 3~mm is comparable to the QS brightness, while at 1.3~mm it is $\sim$1000~K brighter than the QS. Penumbral brightness increases towards the outer boundary in both ALMA bands.
Among the tested umbral models, that of \citet{1994ASIC..433..169S} provides the best fit to the observational data, including both the ALMA data analyzed in this study and data from earlier works. No penumbral model amongst those considered here gives a satisfactory fit to the currently available measurements.
ALMA observations at multiple mm wavelengths can be used for testing existing sunspot models, and serve as an important input to constrain new empirical models.
\end{abstract}


\section{Introduction}

Sunspots are the largest concentrations of magnetic flux on the Sun and a fascinating magnetic phenomenon. Our understanding of sunspots is far from complete in spite of intensive observational and theoretical research over hundreds of years. Whereas the photospheric structure of sunspots has been studied very extensively in the last few decades \citep[e.g.][]{2003A&ARv..11..153S, 2011LRSP....8....4B}, the knowledge of sunspot chromospheres is still relatively poor.

Recent observations with spatial resolution better than 0.5\arcsec\ have revealed that, besides stable large-scale structure, both sunspot umbra and penumbra appear to be organized on small spatial scales, harboring umbral dots embedded in a more uniform and darker background, light bridges, radially elongated penumbral filaments, and penumbral grains \citep[see, e.g.][]{2011LRSP....8....3R, 2009ScChG..52.1670B}. Unfortunately, the large size of sunspots and the dominance of fine-scale dynamic structure make modelling of sunspots as a whole highly intricate. Whereas realistic simulations of complete sunspots do exist \citep{2009ApJ...691..640R, 2011ApJ...740...15R}, they have been restricted to the photosphere. An extension to the chromosphere and corona by \citet{2017ApJ...834...10R} is based on optically thin radiative losses and field aligned heat conduction in the corona, while the chromosphere is treated in local thermodynamic equilibrium (LTE), which is known to be a highly simplified representation of chromospheric radiative transfer.

An important role in deriving the thermal structure of sunspots and its further comparison with theoretically predicted thermal stratifications is played by empirical modelling. Using empirical models spanning the whole solar atmosphere, from beneath the photosphere all the way into the corona, we can obtain information about convective energy transport in umbrae and penumbrae as a function of height, as well as about the layers where mechanical energy transport and deposition become important. There have been numerous attempts to model sunspots (semi-)empirically, based on either strong spectral lines in the visible or on lines in the UV with non-LTE radiative transfer \citep[see][for an overview]{2003A&ARv..11..153S}. Typically sunspot umbrae and penumbrae were modelled separately, with the former getting most of the attention. As a result, the total number of empirical umbral models is large, although not all models are independent of each other \citep{2003A&ARv..11..153S}. Usually single-component umbral models describe the prevalent dark core of the umbra, which is believed to be relatively homogeneous. For instance, Avrett's ''Sunspot sunspot model'' of the umbral photosphere, chromosphere and transition zone, published in 1981, is synthesized from the efforts of a number of modellers \citep{1981phss.conf..235A}. \citet{1986ApJ...306..284M} improved Avrett's model further in the photospheric layers and this model has been the ''standard'' sunspot model for many years. The models evolved further by adding new observational data and new approaches \citep{1994ASIC..433..169S, 2006ApJ...639..441F, 2009ApJ...707..482F} and by employing non-LTE inversions of chromospheric and photospheric lines \citep{2007ApJS..169..439S, 2016ApJ...830L..30D}.

Empirical models of penumbrae are, on the contrary, rare, due to the prominent fibrilar structure of penumbra, which requires in the first place multi-component fine-scale modelling of horizontal structure. All in all, the various available umbral (\& penumbral) models are rather diverse and there is a distinct need to distinguish between them, and validate or rule out them.

As shown in \citet{2014A&A...561A.133L} observations at submm and mm wavelengths can be used as complementary diagnostics of sunspot models at chromospheric heights. Radiation at these wavelengths is formed in LTE and comes from the low to mid chromosphere. However,  prior to the advent of the Atacama Large Millimeter/Submilimeter Array (ALMA) sunspot observations useful for this purpose at these wavelengths were very rare due to the generally insufficient spatial resolution of instruments operating at submm/mm wavelengths. The spatial resolution needed to resolve sunspots and their structure can be achieved by observing at shorter wavelengths, with bigger size dishes, or by employing interferometric observations. \citet{1995ApJ...453..517L} observed several sunspots with the James Clerk Maxwell Telescope (JCMT) at 0.35~mm, 0.85~mm, and 1.2~mm with spatial resolution of 14-17\arcsec, which is higher than the typical resolution of submm/mm single dishes. In the analyzed sunspots, umbrae were significantly cooler than the quiet Sun at submm and remained cool at short mm-$\lambda$, while the penumbrae were brighter than the quiet Sun in the range of observed wavelengths.
Using the Nobeyama 45-m telescope \citet{2015ApJ...804...48I} deduced that the umbral brightness temperature was not higher than the brightness of quiet-Sun regions at 3.5~mm (with a spatial resolution of 19\arcsec) and at 2.6~mm (spatial resolution of 15\arcsec).  White et al. (2006) reported observations of a sunspot with the 10-element Berkeley-Illinois-Maryland Array (BIMA) at 3.5 mm with a resolution of around 10\arcsec, which represents the highest spatial resolution at this wavelength before ALMA became available for solar observations in 2016 \citep{2017SoPh..292...88W,2017SoPh..292...87S}. The umbra could not be clearly resolved in the BIMA images due to the limited resolution, but it was found to be the darkest feature in the interferometric maps, similar to or cooler than the quiet Sun \citep{2014A&A...561A.133L}.

\citet{2014A&A...561A.133L} also showed that models predict sunspot umbrae to be darker than the QS at short mm wavelengths, but brighter at longer wavelengths. How large the contrast is and where the transition occurs, depends strongly on the model. Therefore, there is a strong need for high resolution observations at mm wavelengths that can clearly isolate the umbra and penumbra in order to distinguish between the models.  Such observations are provided by ALMA, which can achieve a sub-arcsecond resolution in the submm/mm range \citep{2010SPIE.7733E..17H}.

Using ALMA Science Verification data \citet{2017ApJ...841L..20I}, hereafter Paper I, report the discovery of a brightness enhancement in the center of a large sunspot umbra at a wavelength of 3 mm, which was observed in the mosaic mode on 2015 December 16. In this paper we extend their analysis by comparing ALMA observations of the same sunspot at two different millimeter wavelengths with the predictions of sunspot umbral and, for the first time, penumbral models.

In Sect.~\ref{obs} we present the results of the ALMA sunspot observations at 1.3 mm and study the statistical brightness distributions for different parts of the sunspot. This wavelength was not considered by \citet{2017ApJ...841L..20I}, who focussed on the umbral brightening at $\lambda$=3~mm. In Sect.~\ref{obs} we also analyze brightness distributions within umbral and penumbral boundaries at 3 mm to study the central enhanced brightness reported in Paper I. The summary of the observational results is presented in Sect.~\ref{sum}. In Sect.~\ref{comp} the brightness at 1.3~mm and 3~mm are compared with the models of umbra and penumbra available in the literature as well as with the other observational data. The results are discussed and conclusions are drawn in Sect.~\ref{dis}.

\begin{figure*}
  \centering
            \includegraphics[width=0.65\textwidth]{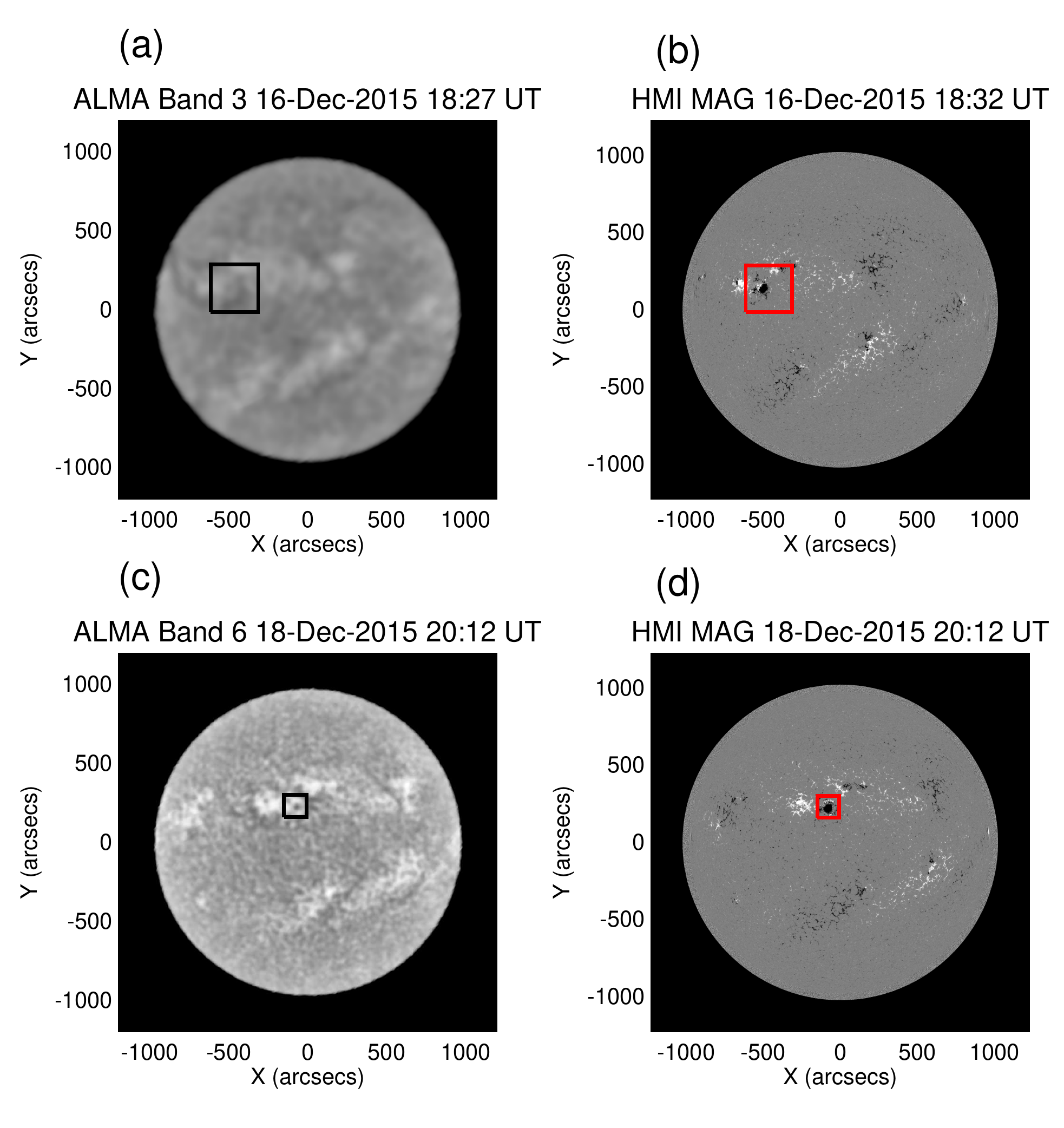}
      \caption{Full-disk ALMA brightness images (a) obtained on 2015 December 16 at 18:32 UT at $\lambda$=3 mm and (c) on 2015 December 18 at 20:12 UT at $\lambda$=1.3 mm. (b) and (d) display the corresponding SDO/HMI LOS magnetograms. The rectangles indicate the positions of ALMA interferometric FOV, including the large sunspot of AR12470.
              }
         \label{fig_fulldisk}
\end{figure*}

\section{ALMA observations and Data Analysis}\label{obs}

The ALMA observations were carried out on 2015 December 16 in Band 3 (100 GHz, corresponding to a wavelength of 3 mm), and on 2015 December 18 in Band 6 (230 GHz, 1.3 mm),   in a compact array configuration which included twenty-two 12m antennae and nine 7m antennae of the Atacama Compact Array (ACA). The full-width at half-maximum (FWHM) of the synthesized beam was 4.9\arcsec $\times$2.2\arcsec\ and 2.4\arcsec $\times$0.9\arcsec\ at 3 mm and 1.3 mm, respectively. The maps were derived from mosaic observations with 149 pointings to cover a field-of-view (FOV) of 300\arcsec $\times$300\arcsec\ at 3 mm and 142.7\arcsec $\times$138.9\arcsec\ at 1.3 mm, respectively. Single-dish full-disk fast scanning was carried out simultaneously, with the FWHM of the primary beam being about 58\arcsec\ and 25\arcsec\ at 3 mm and 1.3 mm, respectively \citep[see][]{2017SoPh..292...88W}. The single dish and interferometric data were further combined in the UV plane via feathering to derive the absolute brightness temperature of the interferometric maps \citep{2017SoPh..292...87S}. Details of the observations and image synthesis can be found in Paper I.

The observed FOVs embraced a part of the active region AR12470, which was located in the eastern hemisphere (N13E30) on December 16, and north of disc center (N15E05) on December 18. Figure 1 shows the full-disc ALMA images at 3 mm and 1.3 mm obtained by single-dish observations together with the co-temporal longitudinal magnetograms from the Helioseismic and Magnetic Imager \citep[HMI:][]{2012SoPh..275..207S} on board the Solar Dynamics Observatory (SDO), with rectangles indicating the position of the ALMA interferometric FOV on 2015 December 16 and December 18. We use right-ascension (R.A.) and declination (Dec.) axes for image display, with coordinates measuring the offset from the solar disk center. Hence, images are rotated by the solar inclination angle (P=9\fdg6) from heliographic coordinates.

\begin{figure*}
  \centering
            \includegraphics[width=0.5\textwidth,angle=90]{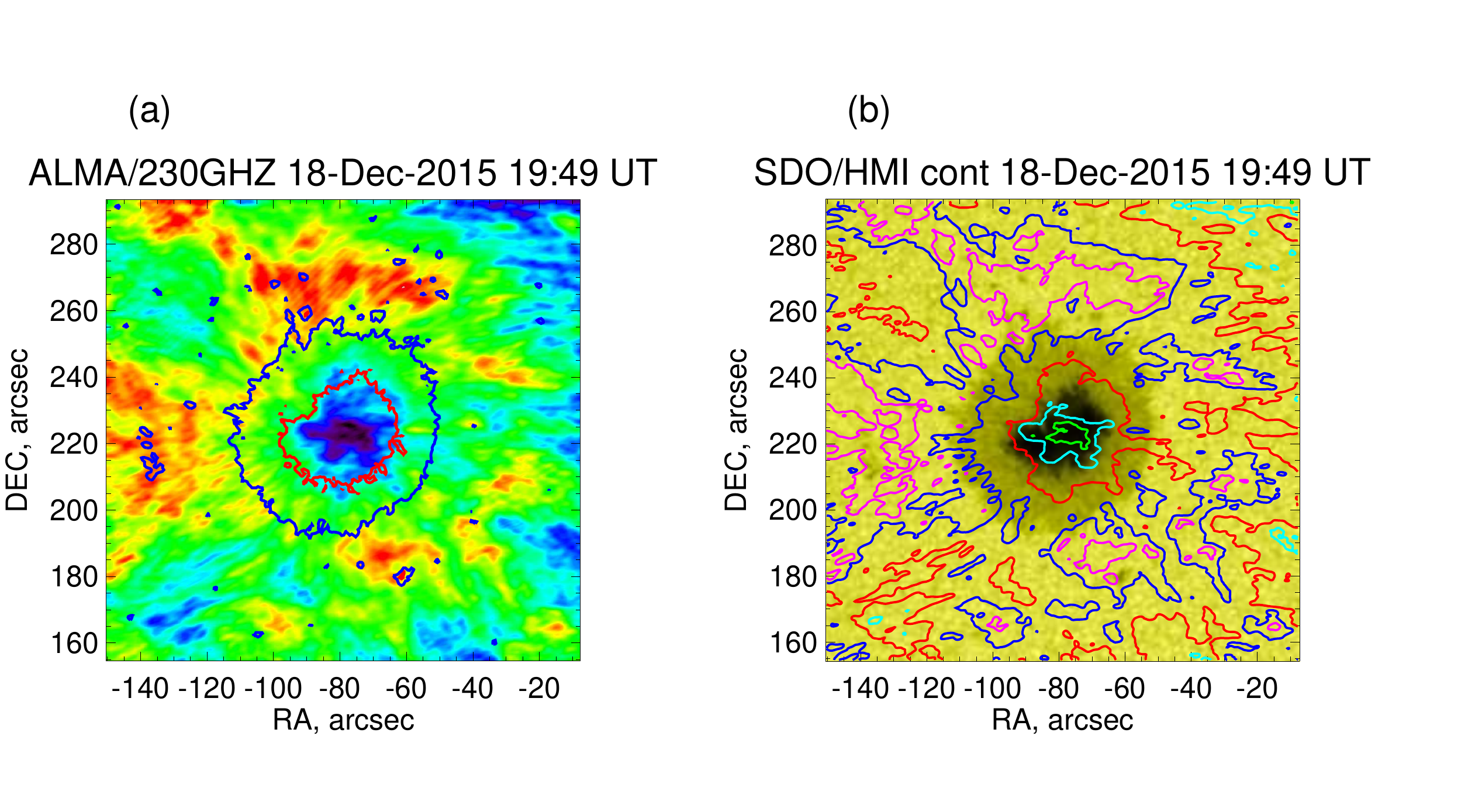}
      \caption{The leading sunspot of AR12470 on 2015 December 18 at 19:49 UT: (a) ALMA 1.3 mm (230 GHz) image with the overlaid umbral (red) and penumbral (blue) contours, (b) SDO/HMI image in the visible continuum. The overlaid color contours indicate 5200 K (green), 5800 K (turquoise), 6500 K (red), 7300 K (blue), and 7800 K (purple) levels in the 1.3 mm image.
              }
         \label{fig_b6_cont}
   \end{figure*}

During the two mapping observations on December 16 and 18 the sunspot region preserved its size and kept a beta-type magnetic structure. While moving toward the central meridian the shape of the sunspot became more symmetric, as can be seen by comparing Fig.~1 from Paper I with Fig.~\ref{fig_b6_cont} of the present paper.


\begin{figure*}
  \centering
            \includegraphics[width=0.85\textwidth,angle=90]{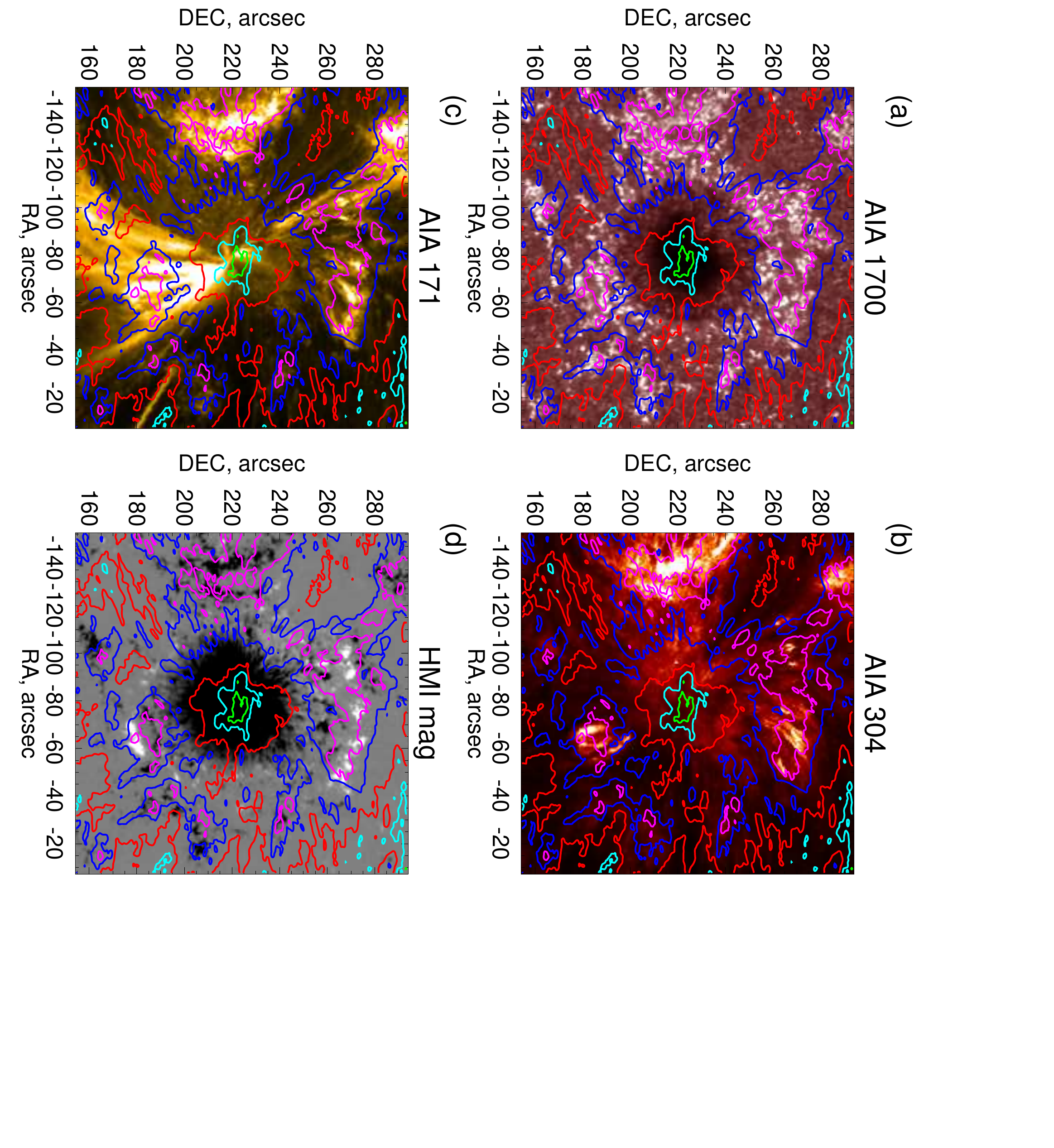}
      \caption{EUV images from SDO/AIA at (a) 1700 Å, (b) He II 304 Å, (c) Fe IX 171, and (d) an SDO/HMI LOS magnetogram in the range [-500,500]G. The overlaid contours are the same as in Fig.~\ref{fig_b6_cont}b. The AIA images are clipped at 50\% of maximum brightness to emphasize the fine structure of brightness enhancements. 
              }
         \label{fig_b6_context}
\end{figure*}


\subsection{Analysis of ALMA 1.3 mm observations recorded on 2015 December 18}\label{obs_b6}

Figure~\ref{fig_b6_cont} shows the ALMA interferometric image obtained at 1.3~mm with the overlaid contours of the umbral (red) and penumbral (blue) boundaries, and a contemporaneous SDO/HMI image in the visible continuum with the overlaid 1.3~mm brightness color contours at 5200 K (green), 5800 K (turquoise), 6500 K (red), 7300 K (blue), and 7800 K (purple). The brightness levels of the red (6500 K) and blue (7300 K) contours very roughly replicate the umbral and penumbral boundaries, respectively. The umbral and penumbral boundaries were derived from the HMI continuum image as, respectively, 0.65 and 0.90 of the surrounding photospheric white-light intensity following \citet{2007A&A...465..291M}. The brightness levels of the green (5200 K) and turquoise (5800 K) contours outline the umbral depression in the mm image. The mm brightness within the umbral contour is considerably lower than in the surrounding penumbra or AR plage. Indeed, the center of the sunspot umbra is the darkest region in the ALMA 1.3~mm map, with brightness temperature below 5000~K. The penumbral region shows a brightness level intermediate between the bright plage and the dark umbra.

In Fig.~\ref{fig_b6_context} the 1.3~mm brightness contours are depicted on top of the SDO/AIA images in a number of passbands, including 1700 \AA, He II 304 \AA, Fe IX 171 \AA, as well as an HMI longitudinal (LOS) magnetogram. A good agreement is seen between the ALMA 1.3 mm image and the AIA 1700 \AA\ image, as well as with the AIA 304 \AA\ image, although the latter two wavelengths arise at totally different heights and temperatures. The blue contours at 7300~K outline the enhanced emission seen outside the penumbra in the 1700 \AA\ image (Fig.~\ref{fig_b6_context}a), while the purple $\lambda$=1.3~mm contours at 7800~K correspond to the patches of enhanced brightness in the 304 \AA\ image (Fig.~\ref{fig_b6_context}b). The correlation of the mm brightness with the 1700 \AA\ brightness is tighter than with the 304 \AA\ brightness, which is consistent with the formation height of mm emission obtained from the analysis of 3D simulations of the chromosphere \citep{2015A&A...575A..15L, 2017A&A...601A..43L}. According to \citet{2017A&A...601A..43L}, emission at 1.3 mm is formed in the lower to middle chromosphere (around 1100 km above the photosphere), which is closer to the formation height of 1700 \AA\ emission (believed to be slightly below the temperature minimum) than that of the 304 \AA\ emission (upper chromosphere and lower transition region).
Enhanced brightness at 1.3 mm typically shows a correspondence with enhanced photospheric magnetic field (Fig.~\ref{fig_b6_context}d) and some correspondence with the coronal 171 \AA\ image (Fig.~\ref{fig_b6_context}c).

There are radial inhomogeneities seen in the structure of both sunspot umbra and penumbra at 1.3~mm. To account for these significant variations of umbral and penumbral intensity, we distinguished between inner and outer umbra as well as between inner and outer penumbra when analyzing the ALMA data, and individually investigated the intensity histograms of pixels lying in each of these locations (plotted in Figs.~\ref{fig_b6_umbra}b, d, f and ~\ref{fig_b6_pen}b, d, f). We define the inner umbra as the region within the red ellipse in Fig.~\ref{fig_b6_umbra}, while the outer umbra is defined to lie between the green ellipse and the white umbral boundary contour in Fig.~\ref{fig_b6_umbra}. The inner penumbra is confined between the white umbral boundary and the blue ellipse in Fig.~\ref{fig_b6_pen}, while the outer penumbra is located between the blue ellipse and the penumbral boundary shown in red (Fig.~\ref{fig_b6_pen}). For comparison we also considered the neighboring part of the AR between the penumbral contour and the black ellipse in Fig.~\ref{fig_b6_pen}.

For each set of pixels we determined typical brightness temperature values as the average of the corresponding set and its root-mean-square (RMS) variation, which are presented in Table~\ref{table1}, together with the brightness at the disk center. The quiet-Sun disk center temperatures of the single-dish images used in this work are around 6000~K and 7400~K, at 1.3~mm and 3~mm, respectively, i.e., slightly larger than the average values recommended for scaling (5900~K and 7300~K) by \citet{2017SoPh..292...88W} but consistent with their 2015 measurements. The mm interferometric images were normalized accordingly, and based on the analysis by \citet{2017SoPh..292...88W} we adopt an uncertainty of order 100 K in the absolute temperatures. The uncertainty in the relative temperatures across the interferometer images is much smaller: \citet{2017SoPh..292...87S} estimate it to be of order 4 K at $\lambda$=3 mm and 10 K at $\lambda$=1.3 mm.

\subsubsection{Umbral Analysis}

\begin{figure}
  \centering
    \includegraphics[width=0.48\textwidth]{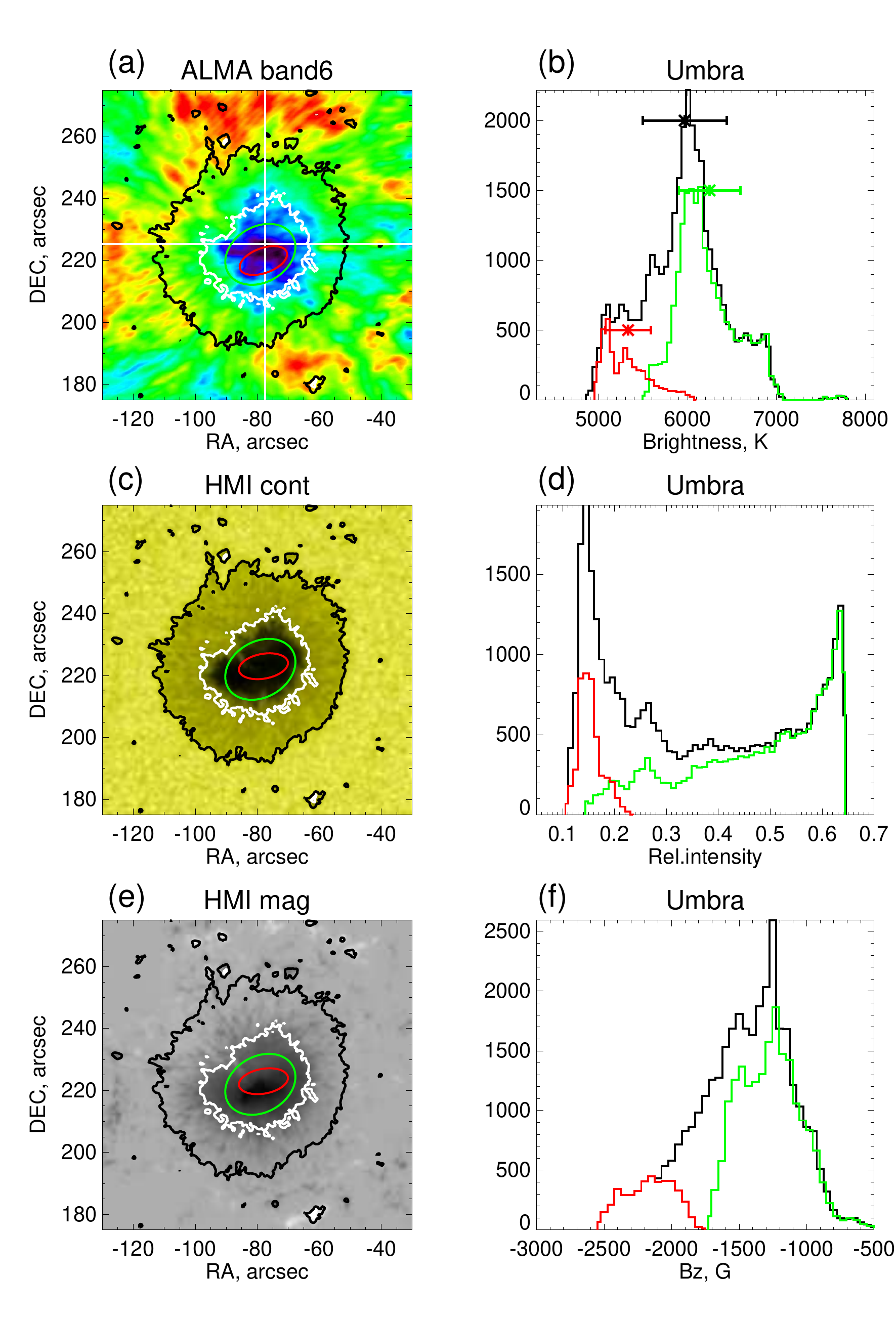}
  \caption{Band 6 umbral analysis: (a), (c), (e) images of ALMA brightness temperature, visible continuum intensity, and LOS magnetic field, respectively. White and black contours indicate the boundaries of the umbra and penumbra. Red and green ellipses outline regions over which the histograms in the right panels were made; (b), (d), (f) Corresponding intensity histograms for 3 different umbral regions. Red histogram: inner umbra within red ellipse, green histogram: outer umbra between green ellipse and white umbral contour, and black histogram: whole umbra within white contour. In the top panel asterisks with error bars indicate mean mm brightness values and their RMS for each region, respectively. White lines in panel (a) indicate the positions of the x- and y-cuts through the minimum of the umbral brightness at 1.3 mm (see Fig.~\ref{fig_profiles}).
              }
         \label{fig_b6_umbra}
\end{figure}

\begin{figure}
  \centering
            \includegraphics[width=0.48\textwidth]{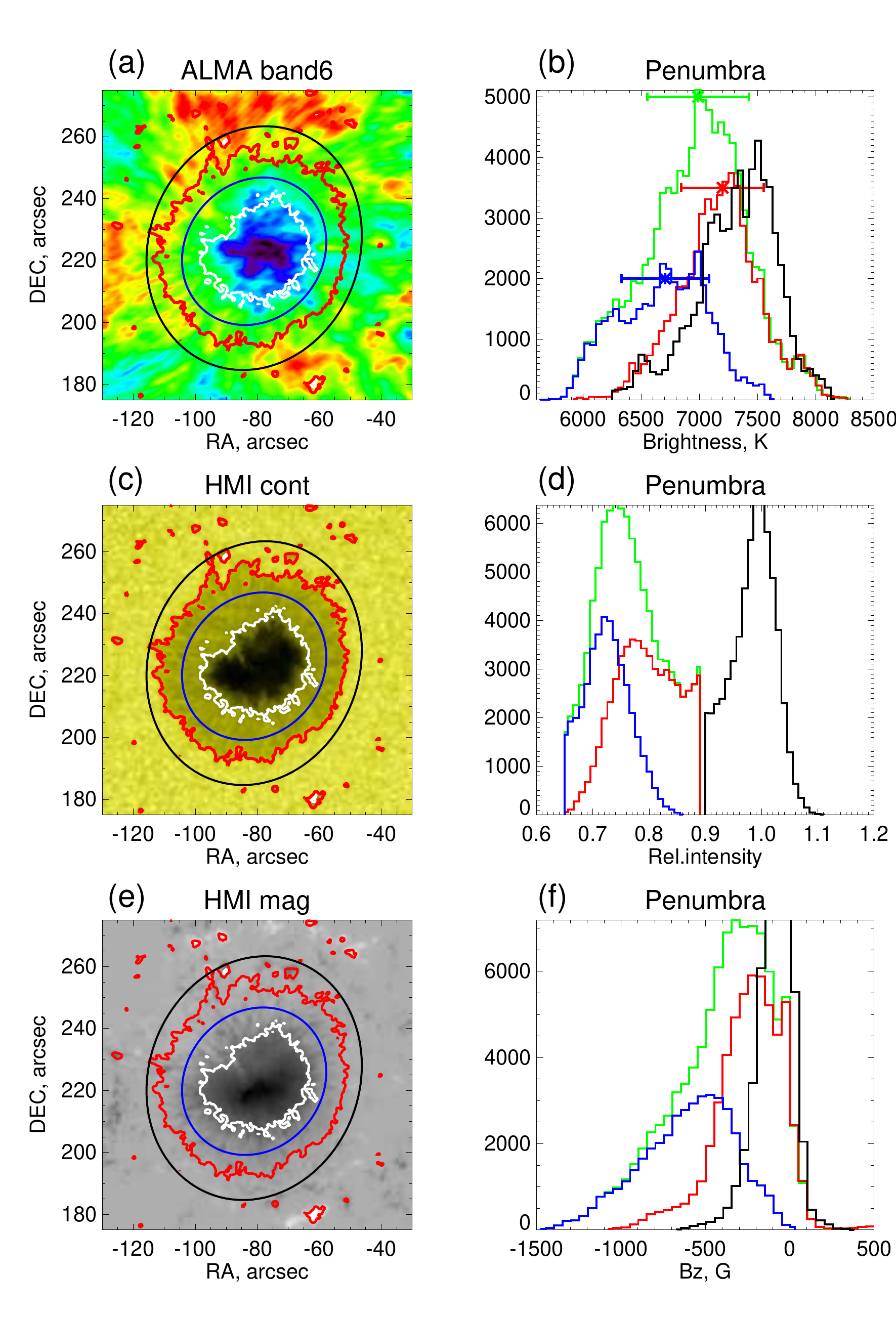}
      \caption{Band 6 penumbral analysis: (a) image of ALMA brightness temperature, (c) visible continuum intensity, and (e) LOS magnetic field, respectively. White and red contours indicate the boundaries of umbra and penumbra, respectively. Blue and black ellipses outline regions over which the histograms in the right panels were made. Panels (b), (d), (f): corresponding intensity histograms for 4 penumbral/super-penumbral regions. The blue histogram is for the inner penumbra between the white umbral contour and the blue ellipse; red histogram: outer penumbra between blue ellipse and red penumbral contour; green histogram: whole penumbra between the red contour and the white umbral boundary; and black histogram: surrounding plage between red penumbral contour and black ellipse. In the top panel asterisks with error bars indicate mean mm brightness values and their RMS for each region, respectively.
              }
         \label{fig_b6_pen}
   \end{figure}

The left panels of Fig.~\ref{fig_b6_umbra} show the images of ALMA brightness temperature in Band 6, in visible continuum intensity, and LOS magnetic field, with the overlaid white and black contours indicating the boundaries of umbra and penumbra, as well as red and green ellipses outlining inner and outer umbral boundaries (chosen by eye as described below). The corresponding intensity histograms of the inner umbra (red), outer umbra (green), and full umbra (black) for 1.3~mm brightness,  white-light intensity, and LOS magnetogram signal, are shown in Fig.~\ref{fig_b6_umbra}b,d,f,  respectively. The inner umbral ellipse, depicted in red, corresponds to the region of the strongest LOS magnetic field, as can be judged from the magnetogram histogram (red curve in Fig.~\ref{fig_b6_umbra}f). Clearly, the inner part of the umbra is the coolest ($\sim$5300 K) feature of the umbra and of the whole sunspot at 1.3 mm (Fig.~\ref{fig_b6_umbra}b). The outer umbra at 1.3 mm is significantly ($\sim$1000 K) brighter and is clearly separated from the inner part in the magnetogram signal and white-light intensity (green histograms in Fig.~\ref{fig_b6_umbra}b,d,f). The shape of the mm intensity histogram of the full umbra, which is characterized by a single peak and almost symmetric tails, differs substantially from the shape of the HMI continuum histogram, which displays two peaks of different intensity at the high and low extremes of the continuum brightness range (Fig.~\ref{fig_b6_umbra}b and d).

 \begin{figure}
  \centering
            \includegraphics[width=0.48\textwidth]{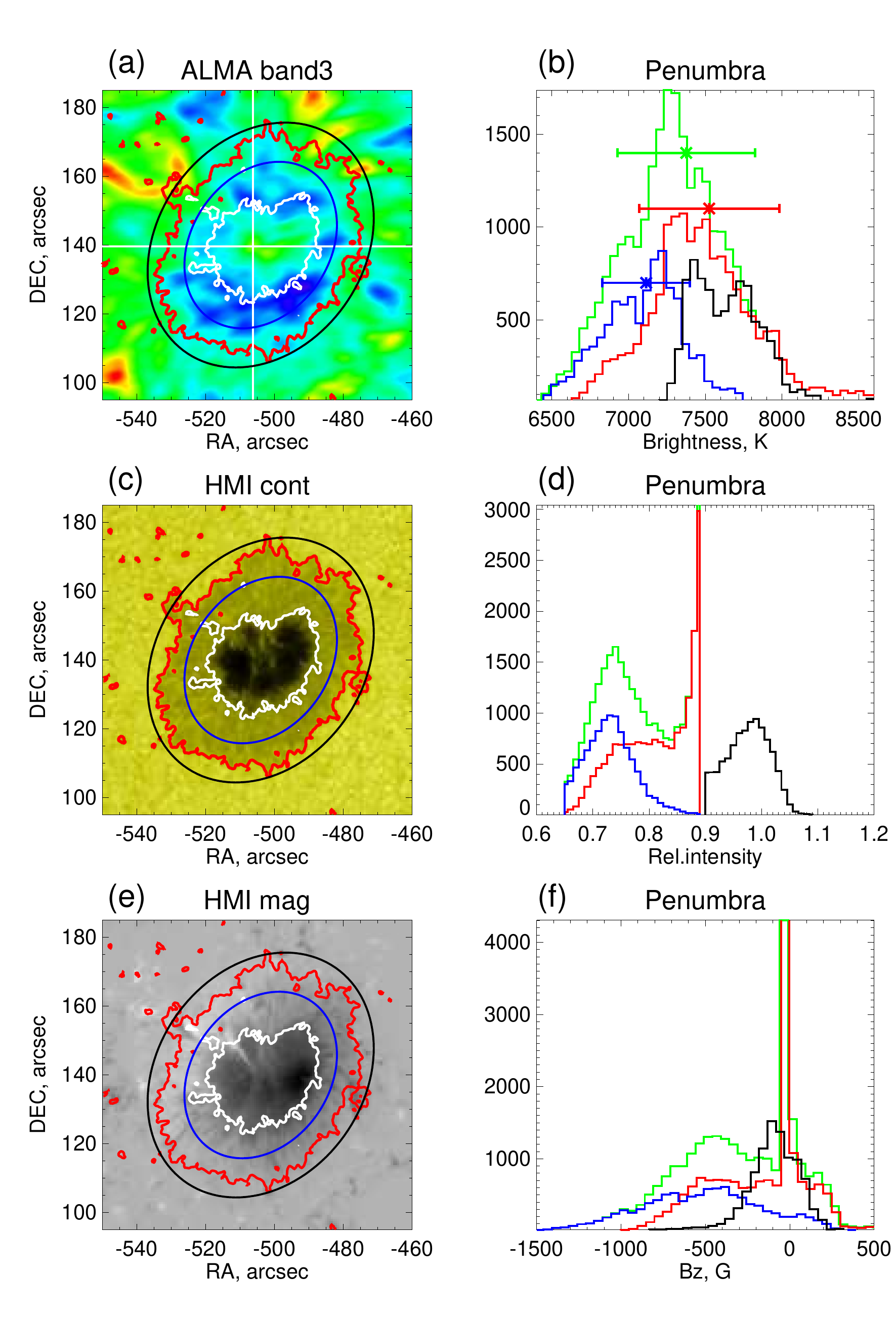}
      \caption{Same as in Fig.~\ref{fig_b6_pen} for the Band 3 penumbral analysis. Vertical and horizontal white lines in panel (a) indicate the positions of the x- and y-cuts through the maximum of the umbral brightness at 3~mm.
              }
         \label{fig_b3_pen}
   \end{figure}

\subsubsection{Penumbral Analysis}

Figure~\ref{fig_b6_pen}a,c,e depict the same three images as in Fig.~\ref{fig_b6_umbra}, but now with overlaid color contours identifying different parts of the penumbra, together with the corresponding intensity histograms plotted in Figs.~\ref{fig_b6_pen}b, d, f. As can be judged from the blue histograms in Fig.~\ref{fig_b6_pen}, the inner penumbra appears rather dark in the 1.3 mm image and comprises the strongest penumbral LOS magnetic field and lowest white-light penumbral intensity. On average the penumbra is $\sim$1000~K hotter than the umbra at $\lambda$=1.3~mm (see Table~\ref{table1}). The outer penumbra (red histograms in Fig.~\ref{fig_b6_pen}) is brighter at 1.3 mm and in white light, and has weaker LOS magnetogram signal. The most striking feature is that at 1.3 mm the difference between the inner and the outer parts of the  penumbra is much larger ($\sim$500~K) than between the outer penumbra and the surrounding plage ($\sim$150~K, see Table~\ref{table1} and black histograms in Fig.~\ref{fig_b6_pen}). This is in agreement with the idea that at lower chromospheric heights the sunspot's magnetic field has expanded significantly beyond the photospheric boundary of the sunspot, so that there is essentially no difference in the physics of the outer penumbra and the adjacent AR plage, which is dominated by sunspot magnetic fields \citep[sometimes called the superpenumbra, e.g.,][]{2003A&ARv..11..153S}.

\subsection{Analysis of ALMA 3 mm observations on 2015 December 16}\label{obs_b3}
\subsubsection{Umbral Analysis}\label{uanal}

The structure of the umbra in ALMA images obtained at $\lambda$=3 mm on 2015 December 16 was analyzed in Paper I, in conjunction with the corresponding images at UV, EUV and visible wavelengths from SDO and the Interface Region Imaging Spectrograph \citep[IRIS:][]{2014SoPh..289.2733D}. Here we summarize the results of that analysis as follows.  The central part of the umbra at $\lambda$=3 mm shows a remarkable brightness enhancement of $\sim$900~K, located close to but not identical with the location of enhanced brightness seen in 1330 \AA\ and 1400 \AA\ images from IRIS (see Figs. 3 and 5 from Paper I). Surprisingly, no clear cospatial counterpart of the mm umbral brightness feature is found in the IRIS data or in AIA images at 1700 and 304\AA, which are otherwise generally similar to the mm image in the large-scale emission from bright plage surrounding the sunspot. Three possible explanations for the observed enhanced umbral radio brightness were proposed in Paper I: the enhancement is an intrinsic property of the umbral chromosphere; a signature of downflowing coronal material interacting with the dense lower atmosphere (as seen in coronal plumes); or produced by dynamic umbral flashes, normally seen in the cores of chromospheric spectral lines. Given the lack of information on the time-dependence of the $\lambda$=3 mm inner umbral brightness, in this paper we adopt the viewpoint that the observed umbral brightness enhancement is an inherent property of the chromosphere above sunspots. We note that the ALMA data presently available are not sufficient to distinguish between the different scenarios, and additional time-resolved mm observations are needed to understand the umbral feature. 

\subsubsection{Penumbral Analysis}

The penumbra at $\lambda$=3 mm shows up as a dark ring surrounding the umbra (Fig.~\ref{fig_b3_pen}). The inner part of the penumbra is $\sim$100~K cooler than the outer part of the umbra (see Table 1). A clear brightness temperature gradient is seen within the penumbra, similar to the one observed at 1.3 mm, with brightness increasing from inside to outside by $\sim$400~K. In addition, the outer penumbra is brighter than its inner part in white light, and has weaker LOS magnetogram signal (Fig.~\ref{fig_b3_pen}d and f). The structure of the outer penumbra at 3~mm is again, as in the 1.3~mm brightness image,  similar to that of the surrounding plage region.  At the same time, the outer penumbra at 3~mm is $\sim$500 K less bright than the inner part of the umbra.

 \subsection{Summary of observational analysis}\label{sum}

 \begin{figure}
  \centering
            \includegraphics[width=0.48\textwidth,angle=90]{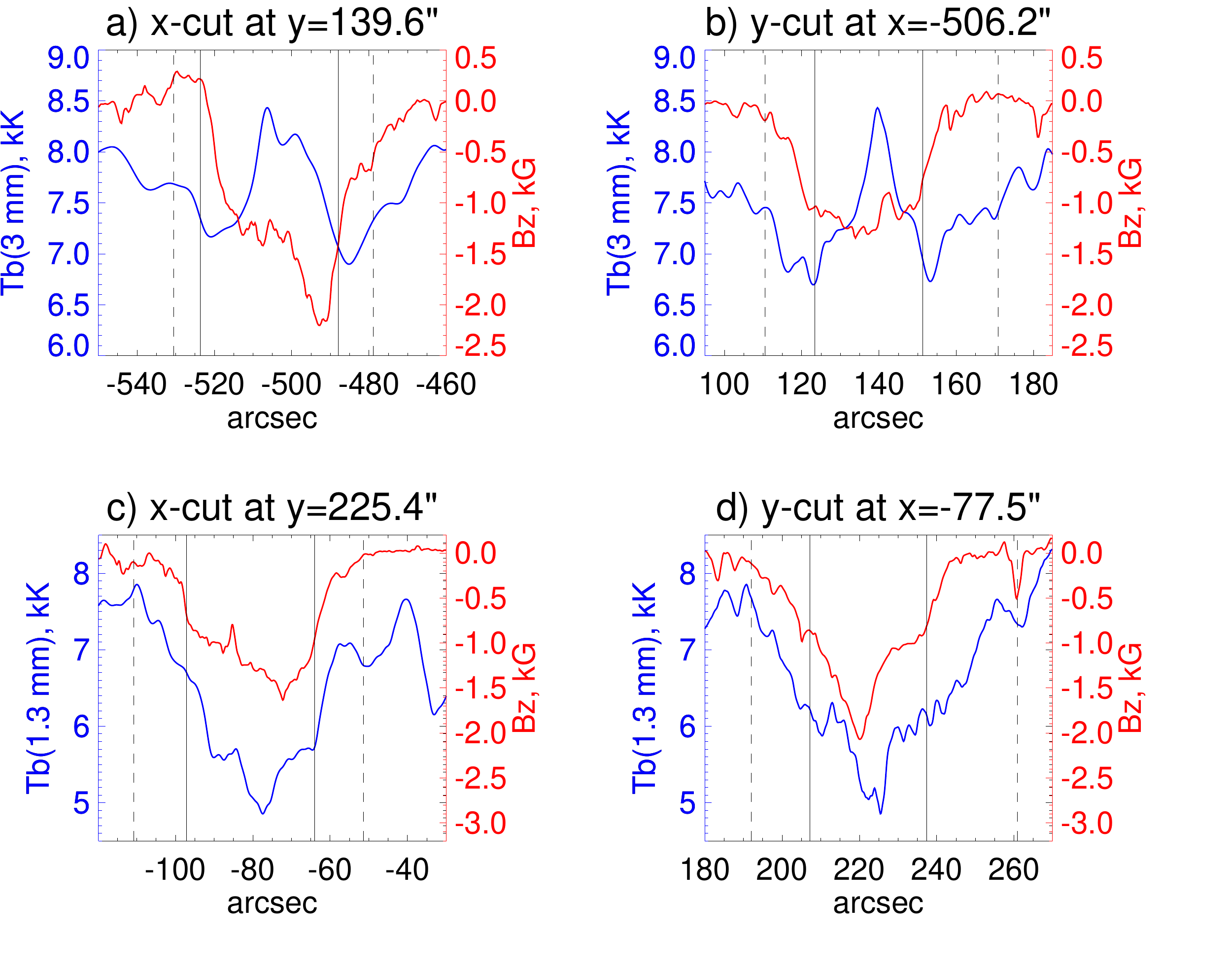}
      \caption{Profiles of the mm brightness (blue, left axis) and of the longitudinal component of the magnetic field (red, right axis) for (a) x-cut at y=139.6”, (b) y-cut at x=-506.2” of the 3 mm image on 2015 December 16, (c) x-cut at y=225.4”, (d) y-cut at x=-77.5” of the 1.3 mm image on 2015 December 18. Solid and dashed black lines indicate the positions of umbral and penumbral boundaries along the cuts, respectively. The positions of the cuts are shown in Figs.~\ref{fig_b6_umbra}a and \ref{fig_b3_pen}a.
              }
         \label{fig_profiles}
   \end{figure}

The summary of the observational analysis is given in Table~\ref{table1}, where we list the mean brightness temperatures and their RMS variations for the umbra and penumbra, analyzed in Figs.~\ref{fig_b6_umbra}, \ref{fig_b6_pen}, \ref{fig_b3_pen}, and in Fig. 5 from Paper I. The last row lists the values for the disk-centered QS brightness derived from the single-dish images in Fig.~\ref{fig_fulldisk}. To summarize the appearance of the sunspot in the two ALMA bands (3~mm and 1.3~mm) we plot in Fig.~\ref{fig_profiles} the profiles of mm brightness along the x- and y-cuts through the brightest umbral pixel at 3~mm and through the darkest umbral pixel at 1.3~mm along with the profiles of the LOS magnetic field.

The distribution of brightness in the umbra at 1.3~mm differs significantly from the distribution at 3~mm (see Table~\ref{table1} and Fig.~\ref{fig_profiles}). The inner umbra is found to be the coolest part of the active region at 1.3~mm, with brightness increasing towards the penumbral boundary. At 3~mm the central part of the umbra shows enhanced brightness, and the brightness drops to inner penumbral values in the outer umbra. Penumbral profiles are similar at 3 and 1.3~mm. The inner part of the penumbra is cooler than its outer boundaries, which are comparable in brightness with the surrounding plage. There are small-scale features, which look similar in the sunspot profiles of the mm brightness and LOS magnetic field, as can be judged from Fig.~\ref{fig_profiles}. The mean values of the LOS magnetic field were $1220\pm400$~G and $1460\pm380$~G in the umbra, and $340\pm300$ G and $380\pm290$~G in the penumbra, on December 16 and December 18, respectively. The similarity of the sunspot structure and of the magnetic field justifies the use of observational results from the two different days jointly when comparing with model brightness predictions.

\begin{table}
\caption{Average brightness temperature $<T_b>$ and its RMS variation $T_b^{\rm  rms}$ in Band 6 and Band 3 for different structures.}             \label{table1}      
\centering                          
\begin{tabular}{ll| c c | c c}        
\hline \hline
\multicolumn{2}{c|}{} & \multicolumn{2}{c|}{1.3 mm (Band 6)} &\multicolumn{2}{c}{3 mm (Band 3)}\\
\multicolumn{2}{c|}{structure}& $<T_b>$, K & $T_b^{\rm  rms}$, K &$<T_b>$, K & $T_b^{\rm  rms}$, K\\
\hline
 \multicolumn{2}{r|}{umbra}   & 5970 &  470 & 7400 & 350\\
  \multicolumn{2}{r|}{inner umbra}  & 5330 & 260 & 7960 & 270\\
  \multicolumn{2}{r|}{outer umbra}  & 6250 & 340 & 7220 & 230 \\
  \multicolumn{2}{r|}{penumbra}  &  6990 & 440 & 7380 &   450 \\
  \multicolumn{2}{r|}{inner penumbra}   & 6700 &  380 & 7110 & 290 \\
  \multicolumn{2}{r|}{outer penumbra}  & 7200 &  350 & 7520 & 460\\
  \multicolumn{2}{r|}{surrounding plage}   & 7340 & 350 & 7740 &  330\\
 \multicolumn{2}{r|}{disk-center QS} & 6000 & -& 7400 & -  \\
  \hline
\end{tabular}
\end{table}

\section{Comparison of ALMA measurements with sunspot atmospheric models}\label{comp}

 \begin{figure*}
  \centering
            \includegraphics[width=0.5\textwidth,angle=90]{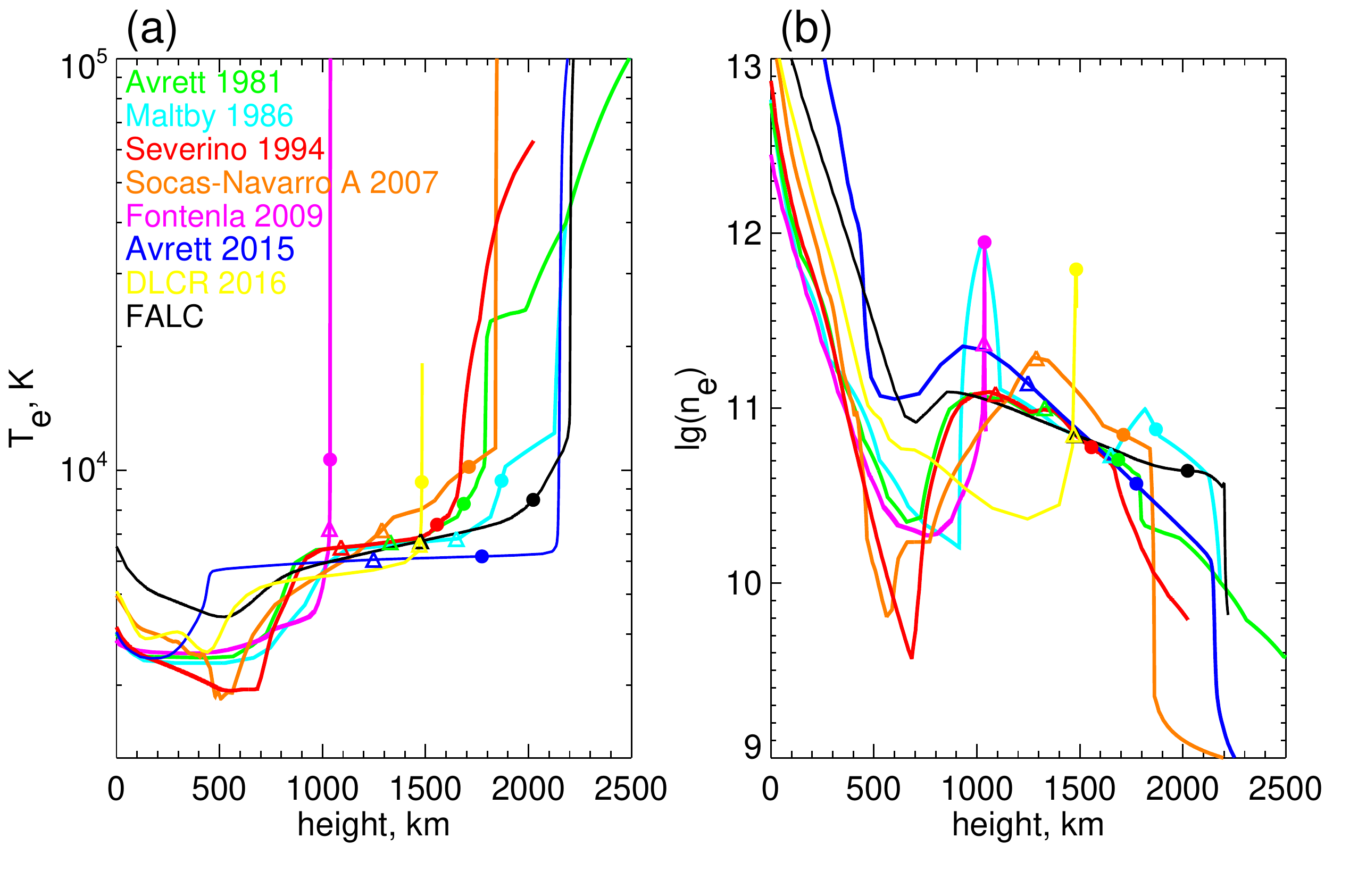}
  \caption{(a) Electron temperature as a function of height in a number of standard models of the solar chromosphere above a sunspot umbra, including the models of \citet{1981phss.conf..235A}, of \citet{1986ApJ...306..284M}, of \citet{1994ASIC..433..169S}, of \citet{2007ApJS..169..439S}, of \citet{2009ApJ...707..482F}, of \citet{2015ApJ...811...87A}, and of \citet{2016ApJ...830L..30D}, marked as DLCR 2016. The individual models are identified by colour as indicated in the figure. The solid black line is the reference quiet-Sun atmosphere FALC. (b) The electron number density as a function of height for the same models as shown in panel (a). Colored triangles and filled circles indicate the effective heights of formation of emission at 1.3~mm and 3~mm, respectively.
              }
         \label{fig_mod_umb}
   \end{figure*}

\begin{figure*}
  \centering
 \includegraphics[width=0.5\textwidth,angle=90]{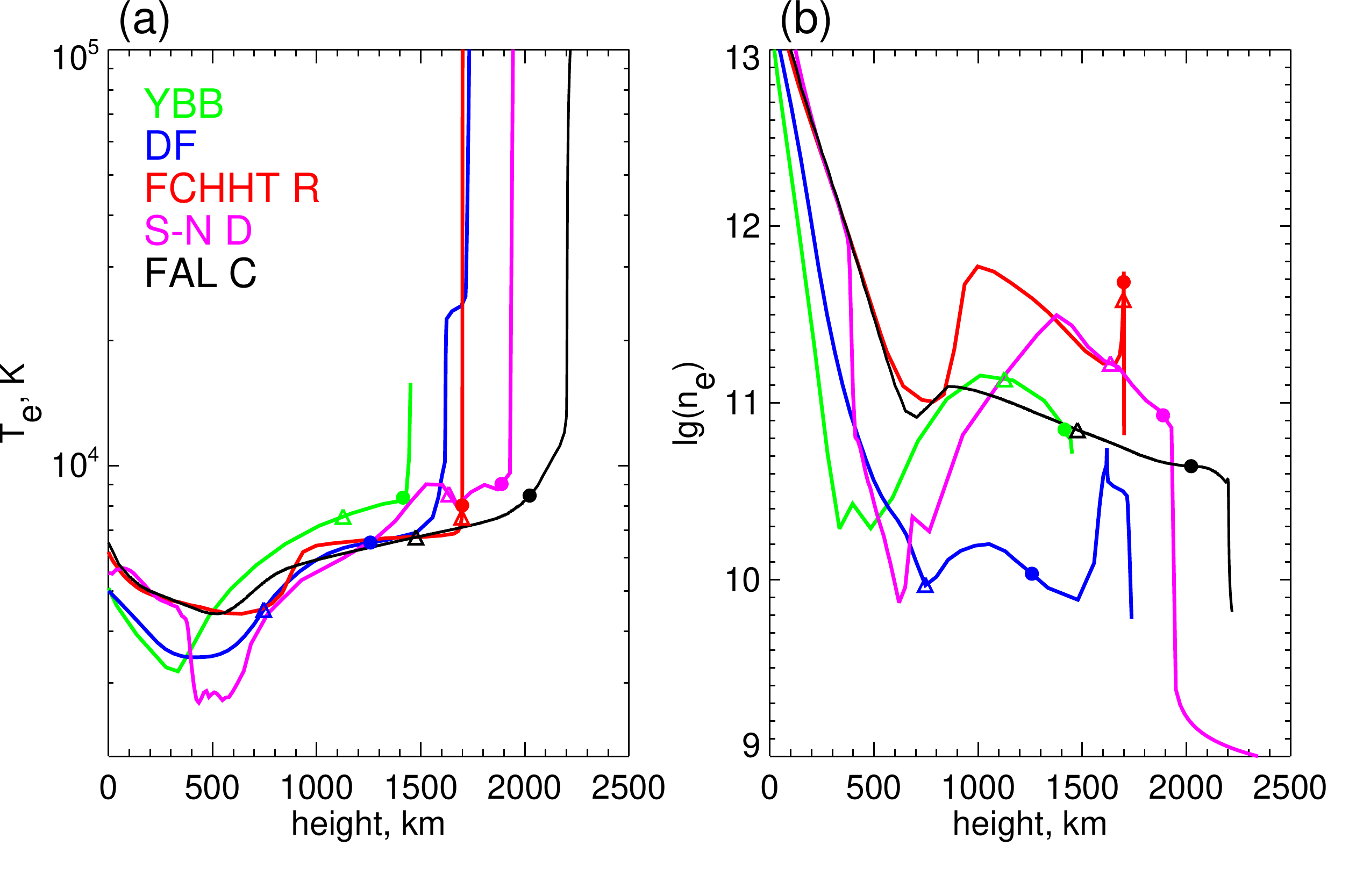}
      \caption{Same as in Fig.~\ref{fig_mod_umb} for the penumbral models of \citet{1984SoPh...92..145Y} (green), \citet{1989A&A...225..204D} (blue), \citet{2009ApJ...707..482F} model R (red), and \citet{2007ApJS..169..439S} model D (violet). Electron temperature and density for the FALC model are plotted in black for comparison. Colored triangles and filled circles indicate the effective heights of formation of emission at 1.3~mm and 3~mm, respectively.}
         \label{fig_mod_pen}
   \end{figure*}

\subsection{Umbral models}\label{umodels}

In this work we build on the same set of umbral models as in \citet{2014A&A...561A.133L}. These include the sunspot model of \citet{1981phss.conf..235A}, model M of \citet{1986ApJ...306..284M}, the sunspot model of \citet{1994ASIC..433..169S}, models A (dark umbra) and B (bright umbra) of \citet{2007ApJS..169..439S}, and sunspot model S of \citet{2009ApJ...707..482F}. The set is supplemented with the recent umbral models of \citet{2015ApJ...811...87A} and of \citet{2016ApJ...830L..30D}. All models, with the exception of that of \citet{2007ApJS..169..439S} and \citet{2016ApJ...830L..30D}, represent the properties of an average sunspot umbra at moderate resolution. The models of \citet{2007ApJS..169..439S} result from the non-LTE inversions of high-resolution spectropolarimetric observations of four Ca {\small II} and Fe {\small I} lines, while the umbral model by \citet{2016ApJ...830L..30D} utilizes the IRIS diagnostics for the same purpose, including Mg~{\small II} h\&k and Mg~{\small II} UV triplet lines and Ni~{\small I} 281.4350 nm line. The electron temperature and electron number density of each of the models are plotted in Fig.~\ref{fig_mod_umb}, together with the reference quiet-Sun atmosphere, which is represented by model C of \citet{1993ApJ...406..319F}, commonly referred to as FALC. More details of the model atmospheres discussed here are given in \citet{2014A&A...561A.133L}. 

\subsection{Penumbral models}\label{pmodels}

In contrast to the numerous umbral models, only a few penumbral models that include the chromosphere have been published. These include those by \citet{1984SoPh...92..145Y}, YBB hereafter, and \citet{1989A&A...225..204D}, DF hereafter, derived from observations of strong chromospheric spectral lines. These models are considered to adequately describe the chromospheric heights but are too cool in the photosphere to reproduce the photospheric observations \citep{2003A&ARv..11..153S}. The third penumbral model is model R (hereafter FCHHT-R) from the set of atmospheric models of \citet{2006ApJ...639..441F, 2009ApJ...707..482F, 2011JGRD..11620108F}, based on the data from \citet{1994ApJ...436..400D} and \citet{1969SoPh....8..275K}, complemented with PSPT observations of the red continuum and Ca II K and with other published data. The final penumbral model we consider, from the non-LTE inversions of \citet{2007ApJS..169..439S}, is model D representing bright penumbra, hereafter labeled S-N D. The height dependence of the electron temperature and electron density for these penumbral models are shown in Fig.~\ref{fig_mod_pen}, together with the FALC model atmosphere.

\subsection{Millimeter-wavelength brightness spectra from umbral models}\label{um}

We have calculated the expected submm/mm brightness temperatures at 32 selected wavelengths in the range 0.1--20 mm for sunspot umbral models listed in Sect.~\ref{umodels}. The calculations were done assuming that the thermal free-free mechanism is responsible for the emission at mm wavelengths. Both types of opacities, H-zero and H-minus, due to interactions between ions and electrons, and between hydrogen atoms and electrons, respectively, were included. The details of the mm brightness calculations can be found in, e.g., \citet{2004A&A...419..747L}.

\begin{figure}
  \centering
            \includegraphics[width=0.45\textwidth]{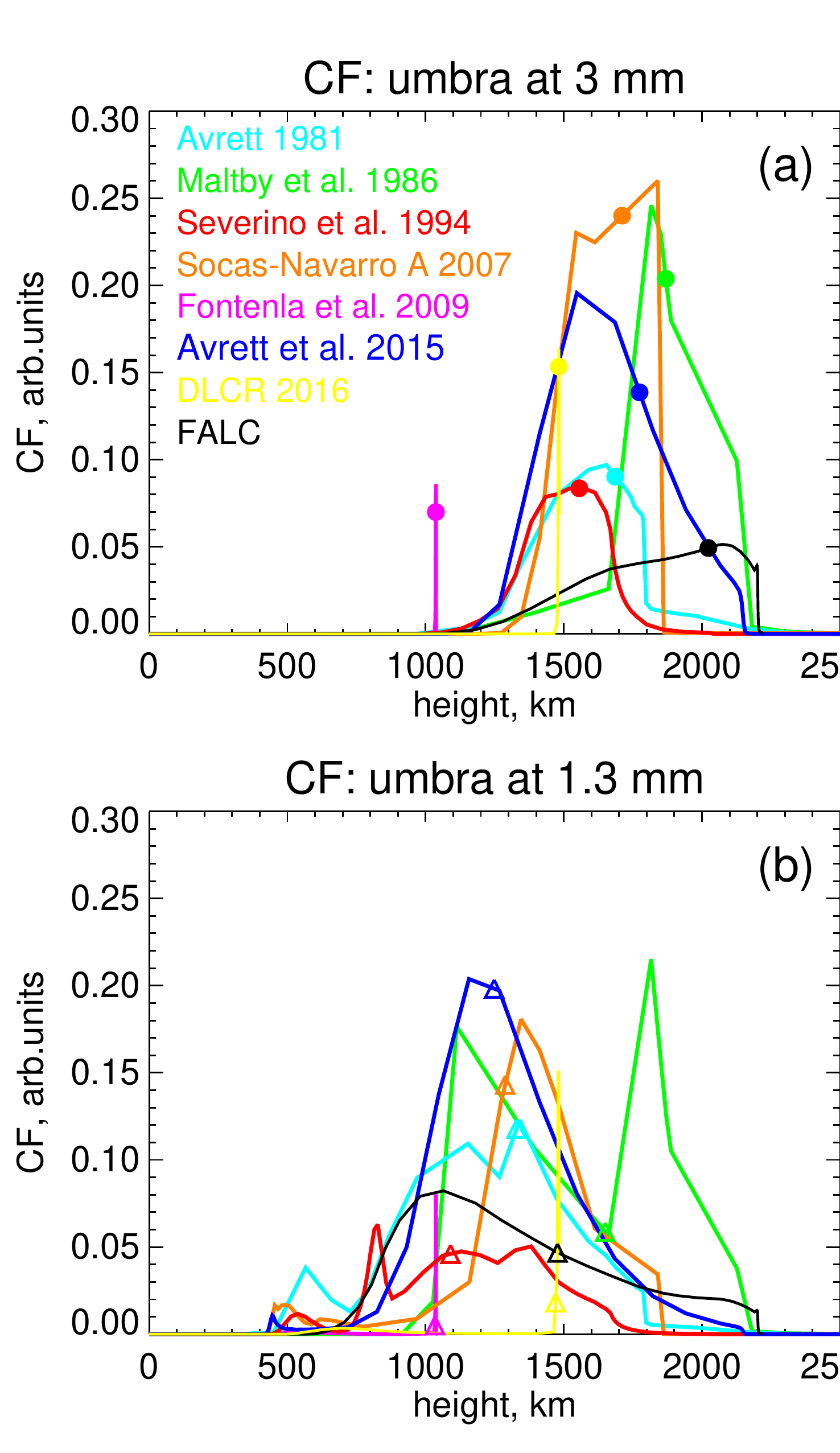}
      \caption{Normalized brightness-temperature contribution functions at 3~mm (a) and at 1.3~mm (b) for the sunspot models depicted in Fig.~\ref{fig_mod_umb}. Colored triangles and filled circles indicate the effective heights of formation of emission at 1.3~mm and 3~mm, respectively.
              }
         \label{fig_mod_umb_cf}
   \end{figure}

The umbral models differ from the FALC model and from each other in the depth and extension of the temperature minimum region and also in the location of the transition region. In Fig.~\ref{fig_mod_umb} the effective formation heights of 1.3~mm and 3~mm emission (also listed in Table~\ref{table2}), marked with the triangles and circles, respectively, are plotted on top of the stratifications of  electron temperature and electron number density for each of the umbral models. Effective formation heights are derived as the heights corresponding to the centroids of the intensity contribution functions (CFs), plotted in Fig.~\ref{fig_mod_umb_cf}, and are indicated in Fig.~\ref{fig_mod_umb} and \ref{fig_mod_umb_cf} with the colored symbols. The colored triangles and circles in Fig.~\ref{fig_mod_umb} and \ref{fig_mod_umb_cf} provide information about the dominant  heights of emission at 1.3 mm and 3 mm, respectively, in the models, while the curves in Fig.~\ref{fig_mod_umb_cf} represent the contribution of various atmospheric layers to this emerging intensity. Millimeter umbral emission at both wavelengths forms over a wide range of chromospheric heights in all the umbral models considered, except for the models of \citet{2009ApJ...707..482F} and of \citet{2016ApJ...830L..30D}. These two models have very sharply defined CFs at both 1.3 mm and 3 mm (violet and yellow curves in Fig.~\ref{fig_mod_umb_cf}), which are very similar to each other at these two wavelengths for each model, because all mm wavelengths become optically thick in the very narrow and rather low-lying transition region present in these models. This TR, located at $\sim$1000~km in the model by \citet{2009ApJ...707..482F} and at $\sim$1500~km in the model of \citet{2016ApJ...830L..30D}, contains a very steep decrease in the electron number density and a strong increase of the electron temperature over a narrow height range. The other models have the transition region at heights similar to the FALC model (about 2000 km above the optical solar surface) and do not possess any abrupt changes with height below that. As a result, a wide range of heights contributes to the emission at 1.3 mm and 3 mm (Fig.~\ref{fig_mod_umb_cf}), while the main contributions at the 2 wavelengths come from heights 200-500 km apart from each other (see Fig.~\ref{fig_mod_umb} and Table 2).

\begin{figure}
  \centering
            \includegraphics[width=0.45\textwidth]{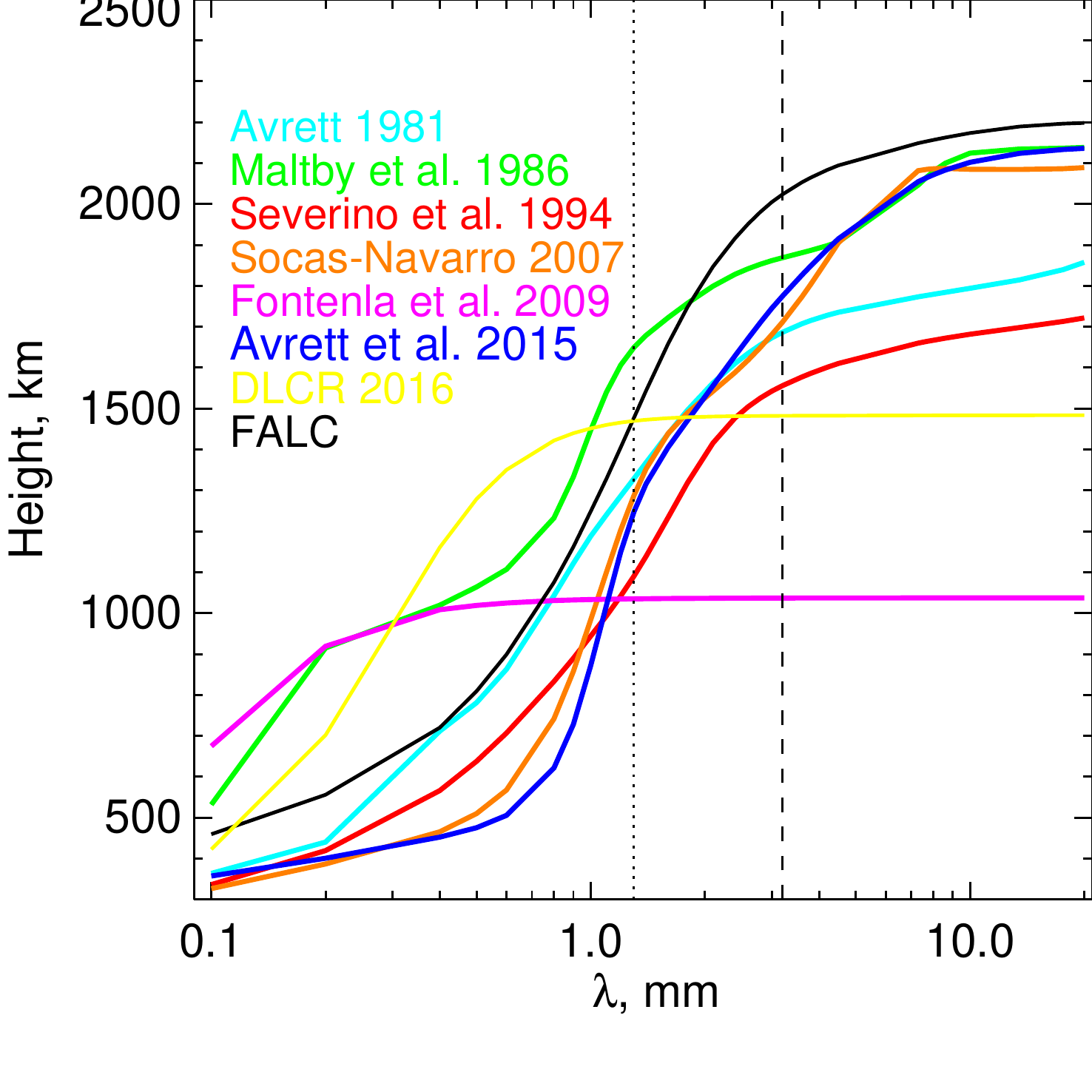}
      \caption{Effective heights of formation of mm emission as a function of wavelength for the same set of umbral models as in Fig.~\ref{fig_mod_umb}. Dotted and dashed lines indicate 1.3 and 3~mm, respectively.
              }
         \label{fig_mod_umb_heff}
   \end{figure}

Effective heights of formation as a function of wavelength for the full range of mm wavelengths and all models investigated are shown in Fig.~\ref{fig_mod_umb_heff}. Again, the largest discrepancies relative to the quiet-Sun FALC model are displayed by the sunspot model of \citet{2016ApJ...830L..30D} and \citet{2009ApJ...707..482F}. The latter model has the largest discrepancy across the whole range of mm wavelengths, with an effective formation height of $\sim$1000~km for all wavelengths longer than $\lambda$=0.5 mm (violet curve in Fig.~\ref{fig_mod_umb_heff}). In the model by \citet{2016ApJ...830L..30D} a constant effective formation height of $\sim$1500~km is reached for wavelengths $>$1~mm (yellow curve in Fig.~\ref{fig_mod_umb_heff}). Except for the \citet{1986ApJ...306..284M}, partly the \citet{2009ApJ...707..482F} and \citet{2016ApJ...830L..30D} models, basically all umbral models find that sub-mm and mm radiation forms at heights lower than in the quiet Sun.

\begin{figure*}
  \centering
            \includegraphics[width=0.65\textwidth,angle=90]{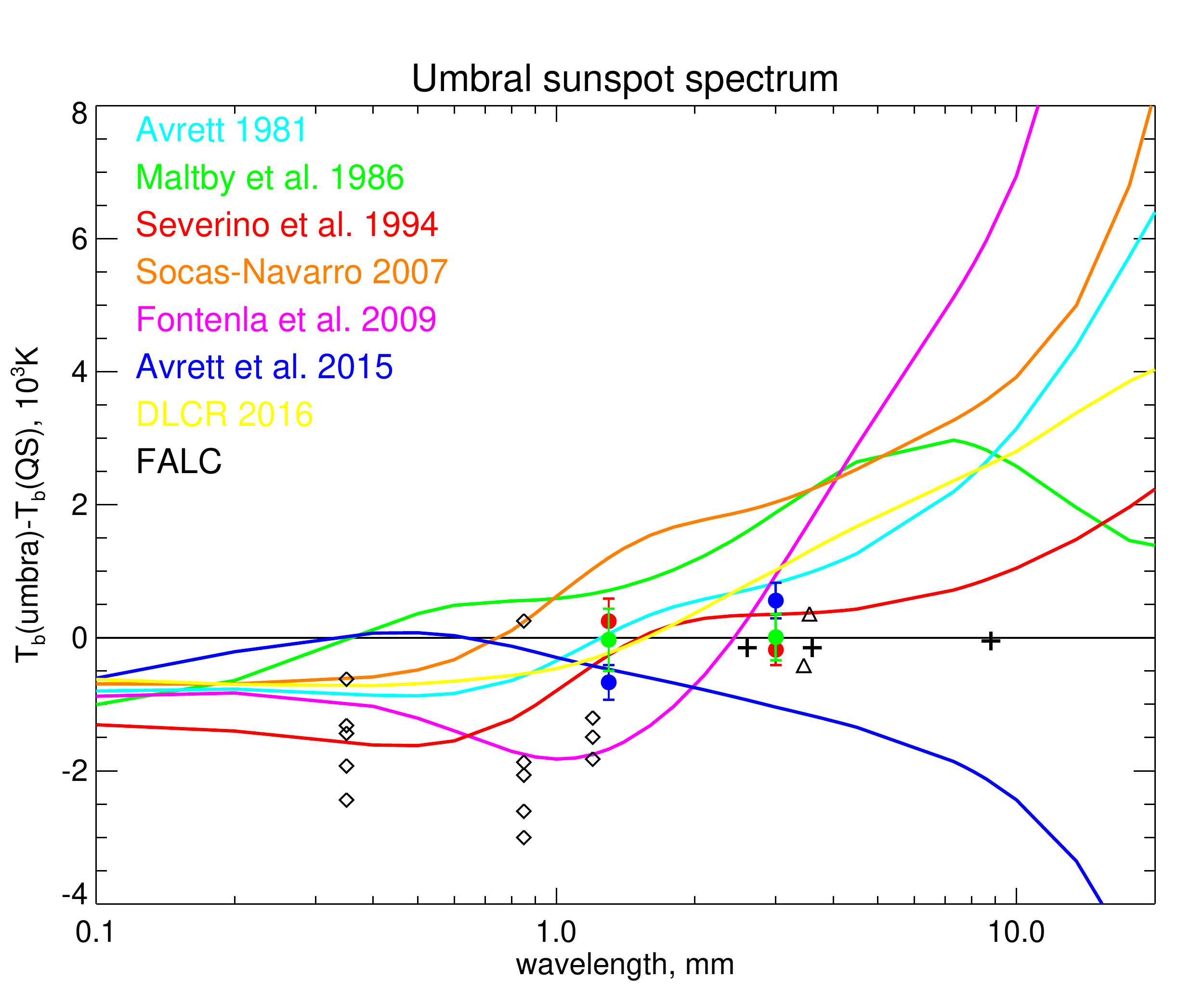}
      \caption{Difference between the umbral brightness (in temperature units) and the QS brightness, plotted as a function of  
      wavelength for the same models as in Fig.~\ref{fig_mod_umb}.  Colored filled circles and error bars indicate the observational mean values together with the RMS values taken from Table~\ref{table1} for the inner umbra (blue), outer umbra (red), and the whole umbra (green) at 1.3~mm and 3~mm. 
      Triangles mark the observational values obtained from BIMA maps at 3.5 mm for the big and small umbrae, respectively. Diamonds stand for the measurements from JCMT at 0.35, 0.85 and 1.2 mm made by \citet{1995ApJ...453..517L}. Pluses indicate the measurements at 2.6 and 3.5 mm from the Nobeyama 45-m telescope obtained by \citet{2015ApJ...804...48I}, and the brightness at 8.8 mm from the NoRH from \citet{2016ApJ...816...91I}.
              }
         \label{fig_mod_umb_tb_rel}
   \end{figure*}

The authors of the umbral models discussed here all chose to use different quiet-Sun models as their reference \citep[see][for details]{2014A&A...561A.133L}. In order to minimize the influence of the reference QS values on the umbral brightness, in Figure~\ref{fig_mod_umb_tb_rel} we plot the difference spectra between the umbral brightness and the QS brightness from the corresponding reference atmospheric models, together with the observational values after subtraction of the disk-centered QS values. The brightnesses at 1.3 mm and 3 mm from ALMA data (this work) are represented by colored filled circles with error bars, which correspond to the observational mean brightness values and their RMS values from Table 1 for the inner umbra (blue), outer umbra (red), and the whole umbra (green). Umbral models typically apply to the central part of an umbra and thus ALMA measurements for the inner umbra (blue circles in Fig.~\ref{fig_mod_umb_tb_rel}) are best suited for comparison with the brightness differences calculated from the models. For completeness, we also plot the observational values obtained from BIMA maps at 3.5 mm by \citet{2014A&A...561A.133L} for the big and small umbrae (triangles), respectively, at a resolution of 12\arcsec; the measurements from JCMT at 0.35, 0.85 and 1.2 mm (diamonds) made by \citet{1995ApJ...453..517L} at a resolution of 14-17\arcsec; brightness observations at 2.6 and 3.5 mm obtained from the Nobeyama 45-m telescope by \citet{2015ApJ...804...48I} at a resolution of 15\arcsec, and at 8.8 mm from the NoRH by \citet{2016ApJ...816...91I} at a resolution of 5-10\arcsec\ (plus signs).

Figure~\ref{fig_mod_umb_tb_rel} shows that, at wavelengths longer than 1.3~mm, the observed umbral brightness, either averaged over the whole umbra (green circles at 1.3~mm and 3~mm), or measured with moderate spatial resolution (at 2.6, 3.5, and 8.8~mm), is not very different from the QS brightness. None of the depicted model curves provides an outstanding fit to the observations at these longer wavelengths. However, the umbral model of \citet{1994ASIC..433..169S}, depicted in Fig.~\ref{fig_mod_umb_tb_rel} by a red curve, shows the best agreement for the ALMA observations of the inner umbrae at 1.3~mm and 3~mm among the models considered in this work. It was already recognized by \citet{2014A&A...561A.133L} as having a good correspondence with the observational data available at that time. 

\begin{table}
\caption{Effective formation heights in km in Bands 6 (1.3 mm) and 3 (3 mm) for different solar atmospheric models.}             \label{table2}      
\centering                          
\begin{tabular}{l| c | c }        
\hline \hline
& 1.3 mm & 3 mm\\
Umbral Model & $h_{eff}$, km &$h_{eff}$, km\\
\hline
Avrett 1981           &    1330 &     1670\\
Maltby et al. 1986           &    1650 &     1860\\
Severino et al. 1994         &    1090 &     1540\\
Socas-Navarro 2007 A   &    1290 &     1680\\
Fontenla et al. 2009 S  &    1030 &     1040\\
Avrett et al. 2015    &    1250 &     1750\\
de la Cruz Rodriguez et al. 2016 & 1470& 1480 \\
  \hline
Penumbral Model & &\\
\hline
Yun et al. 1984 (YBB) &    1130 &     1410\\
Ding \& Fang 1989 (DF) &     750 &     1220\\
Fontenla et al. 2009 (FCHHT R) &    1695 &     1700\\
Socas-Navarro 2007 (S-N D)    &    1640 &     1870\\
  \hline
QS Model & &\\

   \hline
   FALC                  &    1480 &     2000\\
    \hline
\end{tabular}
\end{table}

\subsection{Millimeter-wavelength brightness spectra from penumbral models}\label{mod_penum}

\begin{figure}
  \centering
            \includegraphics[width=0.45\textwidth]{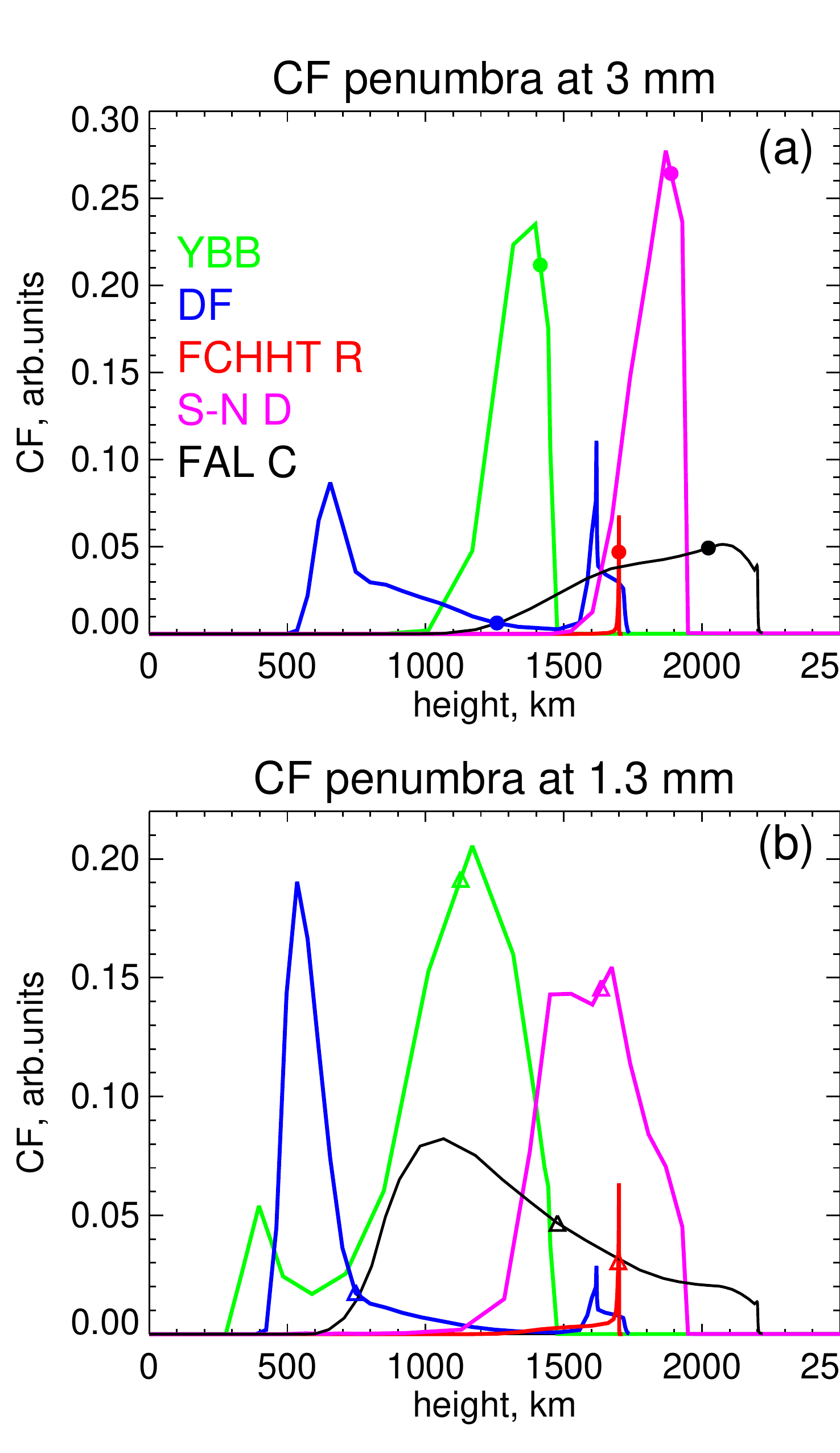}
      \caption{Same as in Fig.~\ref{fig_mod_umb_cf} for the mm CFs from the penumbral models and FALC model. Colored triangles and filled circles indicate the effective heights of formation of emission at 1.3~mm and 3~mm, respectively.
              }
         \label{fig_mod_pen_cf}
   \end{figure}

\begin{figure}
  \centering
            \includegraphics[width=0.45\textwidth]{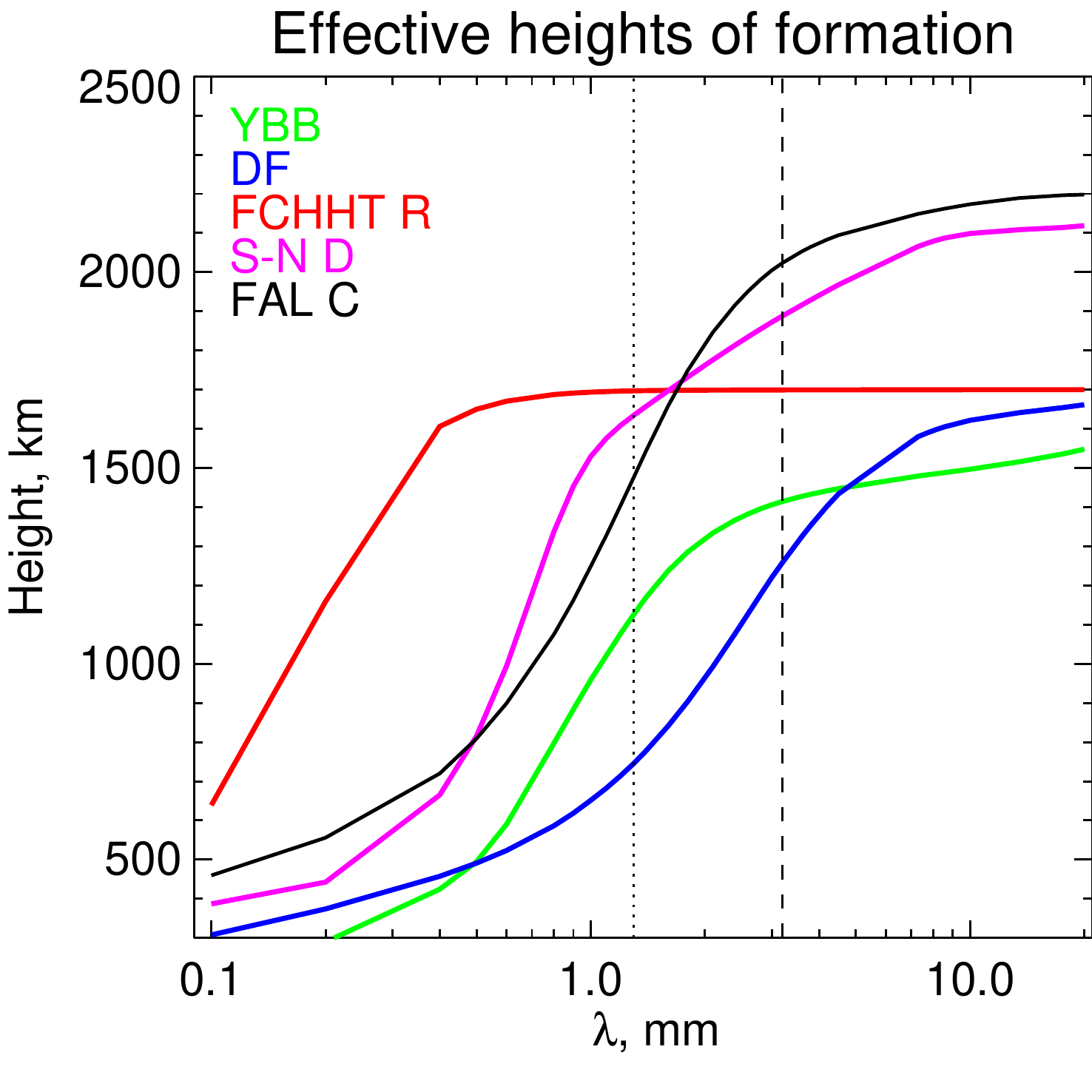}
      \caption{The same as in Fig.~\ref{fig_mod_umb_heff} for the chosen set of penumbral models from Fig.~\ref{fig_mod_pen}. Dotted and dashed lines indicate 1.3 and 3 mm, respectively.              }
         \label{fig_mod_pen_heff}
   \end{figure}

With the exception of FCHHT R, all the penumbral models place the emission at $\lambda$=1.3 and 3 mm at heights that are 300-600 km apart, as can be judged from the colored triangles and circles indicating the effective formation heights at the two wavelengths in Fig.~\ref{fig_mod_pen}, \ref{fig_mod_pen_cf}, and from the heights reported in Table~\ref{table2}. On the other hand, the heights which contribute to the mm emission estimated from different models are significantly different, as is seen from the forms of the CFs, locations of their maxima, and effective formation heights, plotted in Fig.~\ref{fig_mod_pen_cf}. Penumbral emission at both wavelengths forms over a wide range of chromospheric heights in the models considered, except for the model FCHHT R. The dependence of effective formation height on wavelength is shown in Fig.~\ref{fig_mod_pen_heff}. In the YBB (green) and DF (blue) penumbral models, sub-mm and mm radiation forms at heights lower than in the quiet Sun, while in the S-N D (violet) and FCHHT R (red) models the radiation at 1.3~mm is formed higher than in the FALC model (Fig.~\ref{fig_mod_pen_heff}).

\begin{figure*}
  \centering
            \includegraphics[width=0.65\textwidth,angle=90]{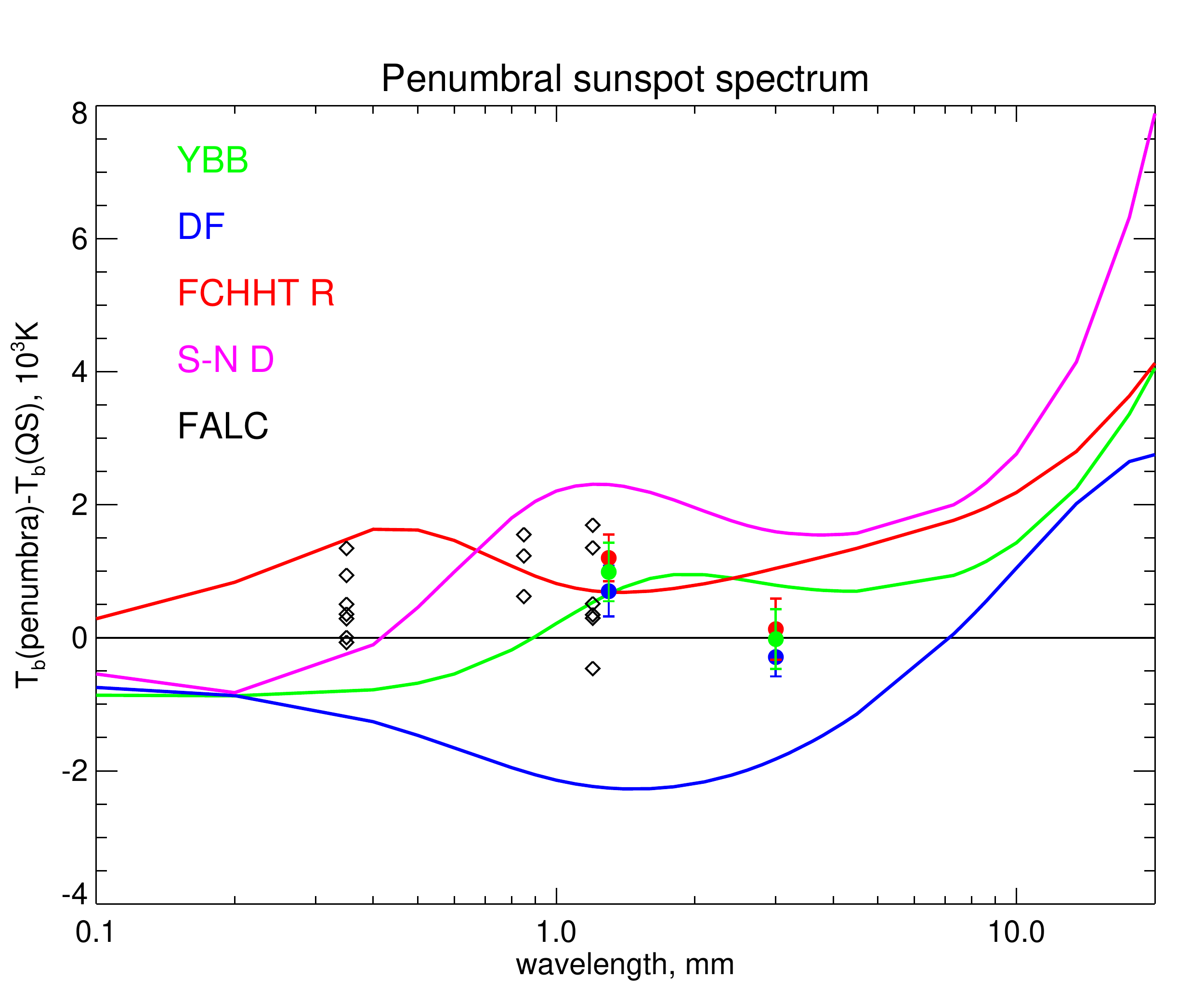}
      \caption{Same as in Fig.~\ref{fig_mod_umb_tb_rel} for the penumbral model spectra. Colored filled circles and error bars indicate the observational penumbral mean values with the RMS values from Table 1 for the inner penumbra (blue), outer penumbra (red), and the whole penumbra (green) at 1.3~mm and 3~mm. 
      Diamonds indicate the brightness measurements at 0.35, 0.85 and 1.2~mm from \citet{1995ApJ...453..517L}.}
         \label{fig_mod_pen_tb_rel}
   \end{figure*}

In Fig.~\ref{fig_mod_pen_tb_rel} we plot the difference between the observed penumbral brightness and the reference QS brightness for each model, distinguishing between inner penumbra (cool, blue), outer penumbra (hotter, red) and penumbra as a whole (green), together with the mm difference brightness spectra, predicted by the models depicted in Fig.~\ref{fig_mod_pen}. As seen from Fig.~\ref{fig_mod_pen_tb_rel} model DF (blue curve in Fig.~\ref{fig_mod_pen}) is too cool to match the observed penumbral mm brightness, while the S-N D (violet) model predicts higher brightness than is observed. Two models, YBB (green) and FCHHT R (red), are in reasonable agreement with the ALMA observations at 1.3 mm, while the penumbral brightness at 3~mm is closer to the QS model brightness (black horizontal line in Fig.~\ref{fig_mod_pen_tb_rel}). The penumbral measurements from JCMT (diamonds in Fig.~\ref{fig_mod_pen_tb_rel}) show a significant scatter and lie between the model brightness spectra of FCHHT R and YBB at the shortest wavelength of 0.35 mm, while at 0.85 mm and 1.2 mm they tend to favor the FCHHT R model. In summary, for the sunspot observed here, the penumbral chromospheric and upper photospheric mm brightness is best reproduced by the FCHHT R and to a slightly lesser extent by the YBB model, which, however, gives a slightly better fit to the ALMA data taken on their own.

\section{Discussion and Conclusions}\label{dis}

The first ALMA observations of sunspots at mm wavelengths, obtained during the solar ALMA SV campaign in 2015, demonstrated that the sunspot umbra exhibits a radically different appearance at 1.3~mm and 3~mm, whereas the penumbral brightness structure is found to be similar at the two wavelengths. The inner part of the umbra is $\sim$600 K brighter than the surrounding QS at 3 mm whereas it is the coolest part of the sunspot at 1.3~mm, being $\sim$700 K cooler than the QS. The appearance of the umbra at 1.3~mm as a dark feature is in agreement with the previous observations of sunspots at submm and short mm wavelengths \citep{1995ApJ...453..517L}. The umbral brightness obtained by \citet{1995ApJ...453..517L} at 1.2 mm is in the range 4600-5200 K, which is also in line with the results reported here at $\lambda$=1.3 mm. Although the synthesized ALMA beam is much smaller than the size of the observed sunspot, the single-dish map which is included in the analyzed interferometric map via the feathering process may suffer from a side lobe effect, especially in the 1.3 mm map \citep{2017SoPh..292...88W}. The ALMA side lobe model for solar observation has not been provided by the ALMA team as yet. Therefore, in the ALMA maps the side lobes of the beam can affect the brightness of a dark region surrounded by bright regions such as seen in the sunspot umbra at 1.3~mm. Therefore, the derived umbral brightness depression in the single-dish map at 1.3~mm should be considered as an upper limit, which suggests that the ALMA umbral brightness values after the side lobe deconvolution would be even closer to those of \citet{1995ApJ...453..517L} at 1.2~mm.

However, the enhanced brightness found in the inner part of the umbra at 3 mm (Paper I) has, to our knowledge, never been reported before. On the contrary, earlier observations provided some evidence for a dark umbra at around 3 mm \citep{2014A&A...561A.133L,2015ApJ...804...48I}. Previous sunspot observations (with BIMA and NoRH) at mm wavelengths were carried out with a resolution of around 10\arcsec, which was insufficient to clearly resolve the umbra within the sunspot. ALMA is the first mm interferometer that offers spatial resolution that not only resolves the umbra but is also sufficient to study some of its fine structure in ALMA's current solar configuration. In future even higher resolution is expected to be achieved.

Comparison of the ALMA Band 6 image with the ultraviolet observations \citep{2017ApJ...845L..19B} shows that the 1.3 mm image is similar to the IRIS image in the Mg~{\small II} h line, which is formed at chromospheric heights. This suggests that there is no contribution from plasma in transition region and corona to the emission at 1.3~mm. On the other hand, \citet{2017SoPh..292...87S} suggests that the plasma in the transition region and corona might contribute to Band 3 (3~mm) images. The large loops apparent in the Band 3 mosaic image, which is included in the solar SV data release, indicate the possible contribution of 0.1~MK plasma because the same loops can be seen in the He~{\small II} 304 \AA\ image obtained with SDO/AIA. At the same time, \citet{2000A&AS..144..169G} provide an estimate of coronal contribution to the chromospheric emission at 3~mm that is only about 1\%. This implies that the contribution of the transition region and corona to the measured brightness temperatures is negligible at both 1.3~mm and 3~mm wavelengths, and they can therefore be directly compared with the chromospheric models.

The ALMA observations of the sunspot have also resolved the penumbral structure at mm wavelengths. At 1.3~mm the penumbra is brighter than the surrounding QS and its brightness increases towards the outer boundary. At 3~mm the inner part of the penumbra is cooler than the QS, but gets brighter towards the outer boundary of the penumbra. In the photosphere at a spatial resolution worse than 1\arcsec, the penumbra loses all its filamentary structure and also looks fairly uniform and is on average considerably brighter than the umbra. Considerable variation in penumbral intensities (from the QS brightness to the brightness of the most intense plage, up to 1000~K in excess of QS) in the range 0.35--1.2 mm was also found by \citet{1995ApJ...453..517L}. It was suggested by those authors that a bright penumbral chromosphere is typical for young complex active regions. To test this hypothesis at longer mm wavelengths, more observations with ALMA of sunspots of different magnetic types and of different ages are vital. The results obtained at mm wavelengths are in contrast to those in the optical range. Thus \citet{2007A&A...465..291M} found that although the umbral brightness varies strongly from one sunspot to another \citep[cf. e.g.][]{2014A&A...565A..52K}, the averaged penumbral brightness changed very little from one sunspot to another. Therefore, there may be a big difference between the behaviour of the penumbra in the photosphere and the chromosphere. However, this must be tested with further observations.

The clear differences found at ALMA wavelengths between the inner and outer umbra, as well as between inner and outer penumbra, are minor compared with the large differences between diverse umbral (and penumbral) models.  The use of different diagnostics to construct the models is likely one of the reasons for the large scatter seen between the models. Alternatively, differences in the models might reflect the diversity of sunspots and differences between them.

Among the tested umbral models that of \citet{1994ASIC..433..169S} provides the best fit to the observational data, both for the ALMA data analyzed in this paper and data from other sources analyzed in earlier works. The values and the slope of the model brightness spectrum are close, although not identical, to the observed brightnesses and their gradient at ALMA wavelengths, which implies that the chromospheric temperature gradient in the model is in reasonable agreement with the ALMA observations. According to this model, the bulk of the emission at 1.3 mm and 3 mm comes from the heights of $\sim$1100~km and 1500~km in the umbral chromosphere, respectively. The chromospheric temperature gradient at these heights is different from that in the QS at the heights of formation of the emission at these wavelengths. The QS emission, estimated from the FALC model, is formed $\sim$500 km higher at both wavelengths.

No penumbral model gives a really satisfactory fit to the currently available measurements. The two models, YBB and FCHHT R, that come closest to being consistent with the data are quite distinct in both their thermal profile and the heights at which the 1.3 mm and 3 mm radiation is emitted. The observed penumbral brightness differs on average only by $\sim$(300--400)~K at 1.3 mm and 3 mm. In terms of temperature stratifications, the YBB model reproduces this difference with a very low temperature gradient, allowing for substantially different heights of formation of emission at the two wavelengths. In the FCHHT R model, however, the two wavelengths are formed very close together, at a height where very rapid temperature increase occurs in the model.

For a definite determination of the temperature gradient in the solar chromosphere at the heights where mm emission is formed, and thus for a precise formulation of the requirements for a successful chromospheric model, additional ALMA sunspot observations are required. Ideally, these would cover multiple spots, to gain better insight into the variations from one sunspot to another, and also more wavelength bands. ALMA observations at multiple mm wavelengths can be used not only for testing existing sunspot models, but can also serve as an important input to constrain new empirical models. We look forward to the use of ALMA Bands 4 and 5, which lie between $\lambda$=3 and 1.3 mm, as well as wavelengths shorter than 1.3 mm for observations of the Sun that can address these issues.

\acknowledgments

This paper makes use of the following ALMA data: ADS/JAO.ALMA$\sharp$2011.0.00020.SV. ALMA is a partnership of ESO (representing its member states), NSF (USA), and NINS (Japan), together with NRC (Canada) and NSC and ASIAA (Taiwan), and KASI (Republic of Korea), in cooperation with the Republic of Chile. The Joint ALMA Observatory is operated by ESO, AUI/NRAO, and NAOJ. The AIA and HMI data are courtesy of the NASA/SDO, as well as the AIA and HMI science teams. M.L. acknowledges NSF grant AST-1312802, NASA grant NNX14AK66G, Russian RFBR grant 16-02-00749, and Saint-Petersburg State University grant 6.37.343.2015. M.S. was supported by JSPS KAKENHI Grant Number JP17K05397. This work has been partially supported by the BK21 plus program through the National Research Foundation (NRF) funded by the Ministry of Education of Korea.

\bibliographystyle{apj}
\bibliography{loukitcheva}

\end{document}